\documentclass[%
 reprint,
superscriptaddress,
 amsmath,amssymb,
 aps,
]{revtex4-2}

\usepackage{graphicx}
\usepackage{dcolumn}
\usepackage{bm}
\usepackage{bbm}
\usepackage{braket}
\usepackage[dvipsnames]{xcolor}

\begin{document}
\preprint{APS/123-QED}

\title{Grover-QAOA for 3-SAT: Quadratic Speedup, Fair-Sampling, and Parameter Clustering}

\author{ Zewen Zhang}
\affiliation{Department of Physics and Astronomy, Rice University, Houston, Texas 77005, USA}
\author{Roger Paredes}
\affiliation{Department of Civil and Environmental Engineering, Rice University, Houston, Texas 77005, USA}
\author{Bhuvanesh Sundar}
\thanks{Current address: Rigetti Computing, Berkeley, California 94710, USA.}
\affiliation{Department of Physics and Astronomy, Rice University, Houston, Texas 77005, USA}
\author{David Quiroga}
\author{ Anastasios Kyrillidis}
\affiliation{Department of Computer Science, Rice University, 
Houston, Texas 77005, USA}
\author{ Leonardo Duenas-Osorio}
\affiliation{Department of Civil and Environmental Engineering, Rice University, Houston, Texas 77005, USA}
\author{ Guido Pagano}
\affiliation{Department of Physics and Astronomy, Rice University, Houston, Texas 77005, USA}
\affiliation{Smalley-Curl Institute, Rice University, Houston, Texas 77005, USA}
\author{ Kaden R. A. Hazzard}
\affiliation{Department of Physics and Astronomy, Rice University, Houston, Texas 77005, USA}
\affiliation{Smalley-Curl Institute, Rice University, Houston, Texas 77005, USA}

\date{\today}

\begin{abstract}
The SAT problem is a prototypical NP-complete problem of fundamental importance in computational complexity theory with many applications in science and engineering; as such, it has long served as an essential benchmark for classical and quantum algorithms. 
This study shows numerical evidence for a quadratic speedup of the Grover Quantum Approximate Optimization Algorithm (G-QAOA) over random sampling for finding all solutions to 3-SAT problems (All-SAT). 
G-QAOA is less resource-intensive and more adaptable for 3-SAT and Max-SAT than Grover's algorithm, and it surpasses conventional QAOA in its ability to sample all solutions. We show these benefits by classical simulations of many-round G-QAOA on thousands of random 3-SAT instances. We also observe G-QAOA advantages on the IonQ Aria quantum computer for small instances, finding that current hardware suffices to determine and sample all solutions. 
Interestingly, a single-angle-pair constraint that uses the same pair of angles at each G-QAOA round  greatly reduces the classical computational overhead of optimizing the G-QAOA angles while preserving its quadratic speedup. We also find parameter clustering of the angles.
The single-angle-pair protocol and parameter clustering significantly reduce obstacles to classical optimization of the G-QAOA angles.
\end{abstract}

\maketitle


\section{Introduction}
Combinatorial optimization problems are of great interest for quantum computing and have been explored using amplitude amplification~\cite{brassard2002quantum}, quantum annealing~\cite{kadowaki1998quantum}, and variational techniques, such as the variational quantum eigensolver (VQE)~\cite{mcclean2016theory} and the quantum approximate optimization algorithm (QAOA)~\cite{Hadfield2019FromAnsatz}. QAOA is suited for combinatorial optimization due to its simple protocol~\cite{farhi2014quantum} and its ability as a hybrid variational algorithm to balance quantum and classical resources to adapt to the constraints of practical quantum computers. 

QAOA has been applied to various combinatorial optimization tasks, with some even demonstrating its capabilities on state-of-the-art quantum devices~\cite{Pagano2020, harrigan2021quantum,Zhu2022Multi-roundComputer,moses2023race,shaydulin2023qaoa,shaydulin2023evidence,lubinski2023optimization,dupont2023quantum,maciejewski2023design}.
Recent explorations on combinatorial problems such as MaxCut, Ising, spin glass, and $k-$SAT focused on the classical optimization of QAOA
-- including employing strategies centered on fixed~\cite{wurtz2021fixed} or smooth angles~\cite{kremenetski2023quantum}, finding more intricate angle structures~\cite{yang2017optimizing,sack2021quantum,brady2021optimal,wurtz2022counterdiabaticity,wu2023adiabaticpassage}, or using empirical optimization strategies~\cite{streif2020training,moussa2022unsupervised,farhi2022quantum,galda2023similarity,shaydulin2023parameter,kapit2023approximability,yu2023solution}. 

Among these problems, the Boolean satisfiability problem (SAT) is central for its practical importance. It is an NP-complete problem, encapsulating the challenges inherent to all NP problems~\cite{cook2023complexity,karp2010reducibility, lucas2014ising}. SAT problems have applications spanning fields such as hardware and software design~\cite{khurshid2004testera,Vizel2015BooleanChecking,Dutra2018EfficientTesting,gaber2020computation}, cryptography~\cite{lafitte2014applications}, bioinformatics~\cite{lynce2006efficient}, network reliability~\cite{paredes2019principled} and more~\cite{marques2008practical}.  $k$-SAT problems are defined in terms of Boolean expressions $\phi(m,n)$ consisting of $m$ conjoined clauses, each of which is a disjunction of $k$ Boolean variables, over a set of $n$ Boolean variables. The density, defined as the ratio $d=m/n$, is an important characteristic that influences the hardness of the problem. 

Numerous SAT applications require fair sampling of solutions, namely the capability for SAT solvers to identify comprehensive solution sets rather than just one, a problem known as All-SAT~\cite{yu2014all} and related to model counting~\cite{valiant1979complexity}. 
There has been growing interest in utilizing quantum algorithms to address SAT problems, leveraging techniques like quantum annealing~\cite{battaglia2005optimization,azinovic2017assessment,ayanzadeh2020reinforcement} and amplitude amplification~\cite{cheng2007quantum,alasow2022quantum, varmantchaonala2023quantum}. 
While quantum annealing-based methods do not consistently ensure speedup, amplitude amplification protocols based on the Grover algorithm~\cite{grover} guarantee quadratic speedup over random guessing. 
However, methods rooted in the amplitude amplification necessitate implementing an oracle, demanding ancilla qubits proportionate to the total clause count~\cite{yang2009solution} or increasing the circuit depth (see Appendix~\ref{append:ga}). Alternative methods that reduce the number of ancillas or depth for the oracle are desirable.

In this work, we adopt a recently proposed QAOA variant (G-QAOA) in which the mixer operation, usually implemented with a global qubit rotation, is replaced by a Grover mixer that guarantees the uniform sampling of the solution space~\cite{sundar2019quantum,Bartschi2020GroverPreparation,Zhu2022Multi-roundComputer}. This has been recently applied to SAT problems~\cite{golden2023quantum,mandl2023amplitude} to find solutions, but its scaling with problem size has not yet been studied. 

Here, we present compelling evidence of the quadratic speedup over random guessing of G-QAOA for finding all solutions to 3-SAT problems. We analyze thousands of random 3-SAT instances up to 26 Boolean variables with densities spanning from 2 to 8, encompassing critical thresholds~\cite{kirkpatrick1994critical, mezard2002analytic}. 
Unlike previous classical optimization schemes that simultaneously determine angles for all QAOA rounds, we show that maintaining a single angle pair throughout all G-QAOA rounds retains a quadratic speedup over random guessing. This approach dramatically reduces the classical optimization cost. These angle pairs cluster in a small parameter range across all 3-SAT instances.

This single angle pair approach obtains a quadratic speedup not only for finding solutions of 3-SAT but also for finding a configuration of the problem variables that satisfies as many clauses as possible, referred to as ``Max-SAT"~\cite{asano2002improved,nannicini2019performance}. Max-SAT is particularly important at densities higher than the so-called critical density when typically no solution to the 3-SAT problems exists~\cite{Mezard2002RandomAlgorithm}. Max-SAT is a significant combinatorial optimization problem in its own right, with applications in areas such as system design~\cite{chen2009spatial,dimitrova2018maximum}, machine learning~\cite{malioutov2018mlic,berg2019applications} and bioinformatics~\cite{guerra2012reasoning,martins2017solving}.  

Finally, we find that G-QAOA guarantees a 10-fold fewer qubits and circuit depth at the critical density compared to the standard Grover algorithm. This advantage increases proportionally to the density. Thus, reducing the required quantum resources allows us to run this algorithm on state-of-the-art IonQ quantum computers, corroborating the fair-sampling attribute of G-QAOA.

The paper is structured as follows:
Sec.~\ref{sec:preliminary} delineates the 3-SAT problems under examination and describes the G-QAOA methodologies implemented. Sec.~\ref{sec:result} numerically demonstrates the quadratic speedup of G-QAOA on 3-SAT problems and presents evidence that the single-angle-pair approach maintains the speedup. Sec.~\ref{sec:experiment} shows results from IonQ's Aria machines, and Sec.~\ref{sec:conclusion} outlines this work's conclusions and outlook.

\section{Background: 3-SAT and G-QAOA}
\label{sec:preliminary}

\subsection{3-SAT} 
\label{sec:3sat}
The 3-SAT problem refers to the particular case of Boolean formulas in \textit{conjunctive normal form}~(CNF) in which every clause $i$ has three literals $l_{i_1}, l_{i_2},l_{i_3}$ (see Appendix~\ref{appendix:sat}). This problem has long been a standard combinatorial setting for algorithmic and theoretical studies.
The density $d$ is a crucial indicator characterizing the solution space and its computational hardness. For a 3-SAT instance with small $d$, the problem is under-constrained and likely to be satisfiable, while at large $d$, it is over-constrained and likely to be unsatisfiable.
The computational difficulty for a broad class of state-of-the-art classical algorithms is maximized at a critical density $d_c\approx4.26$~\cite{Mezard2002RandomAlgorithm}.

Given a 3-SAT CNF instance $\phi$, the problem can be translated into a qubit representation through a problem Hamiltonian $H_P$~\cite{lucas2014ising} 
\begin{equation*}
    H_P = \sum_{i=1}^m H_C^{(i)},
\end{equation*}
where each local Hamiltonian $H_C^{(i)}$ is associated to the $i^{th}$ clause $(l_{i_1}\vee l_{i_2}\vee l_{i_3})$ of the instance  and defined as
\begin{equation}\label{eq:clauseH}
    H_C^{(i)} \equiv \tfrac{1}{8} \left(\mathbbm{1} - s_{i_1} {Z}_{i_1} \right) \left(\mathbbm{1} - s_{i_2}{Z}_{i_2}\right) \left(\mathbbm{1} - s_{i_3} {Z}_{i_3}\right),
\end{equation}
where  $\mathbbm{1}_i$ is an identity operator on the $i^{th}$ qubit, ${Z}_j$ is the Pauli operator ${Z}=\ket{0}\bra{0} - \ket{1}\bra{1}$ acting on the $j^{th}$ qubit, and $s_{i_j}\in\{\pm1\}$ denotes the sign of the literal $l_{i_j}$ ($-1$ if negated, and $+1$ otherwise). Therefore, given an assignment of the Boolean variables $\tau\in \{0,1\}^{\otimes n}$ corresponding to a quantum state $\vert \tau \rangle$,  the local energy $E_i$ of the $i^{th}$ clause is
\begin{equation*}
E_i=\langle \tau \vert H_C^{(i)} \vert  \tau \rangle = \begin{cases}
0, & \text{if $\tau$  satisfies $(l_{i_1}\vee l_{i_2}\vee l_{i_3})$},\\
1, & \text{otherwise}.\\
\end{cases}
\end{equation*}
Thus, the expectation value of $H_P(\tau)$ equals the number of unsatisfied clauses of the instance $\phi$ under assignment $\tau$. If $\ket{\tau}$ is a zero-energy ground state of $H_P$, then $\tau$ is a solution of $\phi$.

\subsection{Grover quantum approximate optimization algorithm}

\begin{figure*}[t]%
\centering
\includegraphics[width=0.99\textwidth]{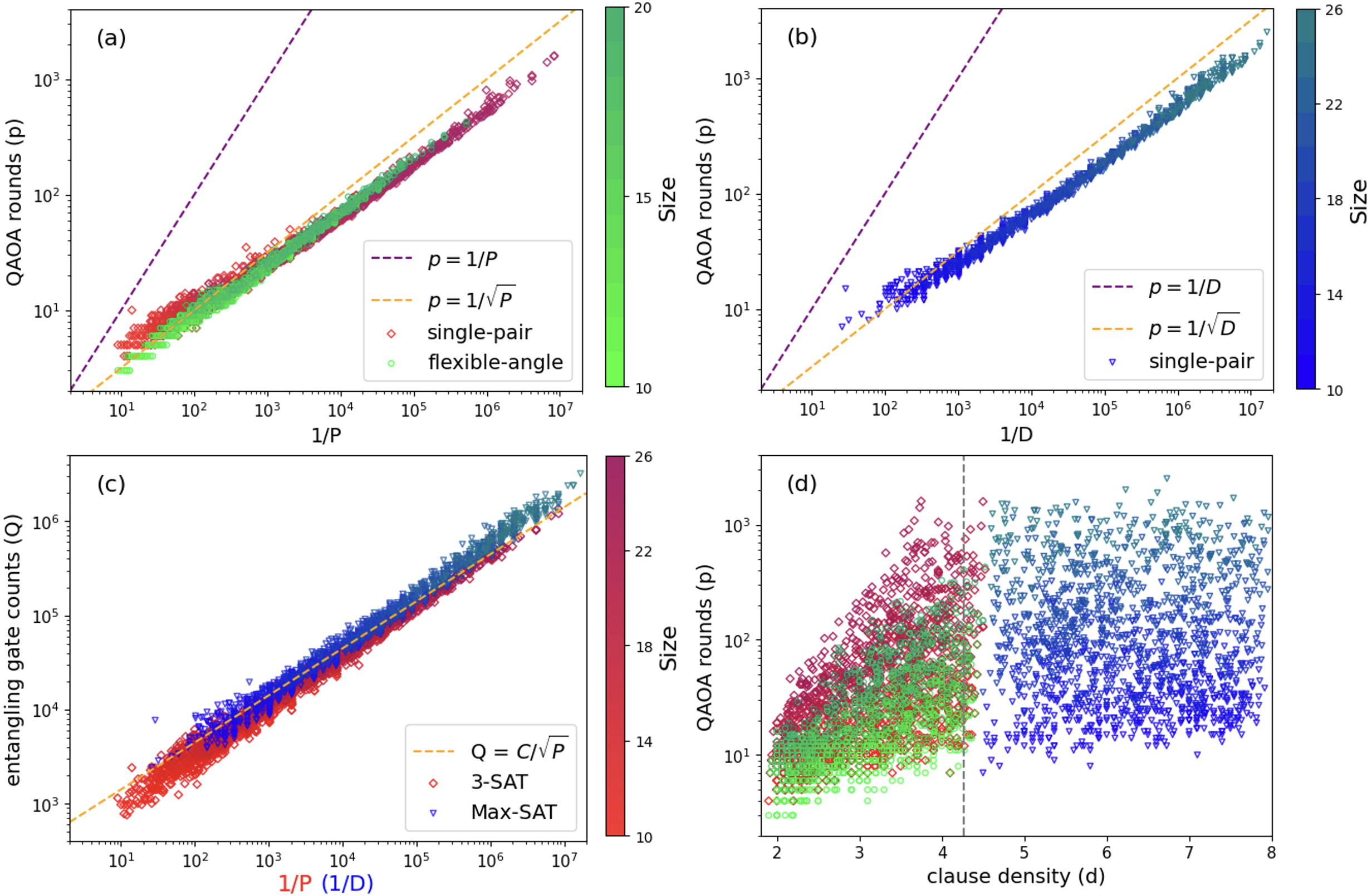}
\caption{\label{fig: speedup} {(color online) \bf Evidence for quadratic speed-up:}
{\bf (a)} \textbf{satisfiable instances}: Minimum G-QAOA rounds $p$ required to achieve 50\% solution probability versus the inverse of the solution probability $P$ 
for $d\in(2,4.5)$. G-QAOA and single-angle-pair G-QAOA (green circles and red diamonds,respectively) are plotted, with the purple line corresponding to $p={1/P}$ and the orange line to 
$p={1/\sqrt{P}}$. Variations in colors differentiate problem sizes. 
{\bf (b)} \textbf{Unsatisfiable instances:} The minimum single-angle-pair G-QAOA rounds vs. the inverse of the Max-SAT probability $D$ 
for $d\in(4.6,8)$. 
\textbf{(c) Evidence for quadratic speedup in terms of required quantum resources.} Number of entangling gates ($Q$) as a function of inverse solution probability $1/P$ (or $1/D$). Red diamonds and blue triangles have the same meaning as in (a), (b). 
{\bf (d) G-QAOA rounds as a function of density $d$:}
green circles, red diamonds, and blue triangles have the same meaning as in (a) and (b). Dashed line marks the critical density elaborated upon in Sec.~\ref{sec:3sat}.
}
\end{figure*}

This section first reviews the G-QAOA with arbitrary variational angles. Then, it discusses the variant we introduce in this paper, the single-angle-pair G-QAOA, where the angles are equal across all G-QAOA rounds.

\medskip
\noindent {\bf G-QAOA.}
For the problems we consider, G-QAOA alternately applies two sets of parameterized unitaries generated by $H_P$ and a mixing operator,
\begin{equation}
\label{eq:Grover_Mx}
    H_M=    \prod_{i=1}^n (\mathbbm{1}_i + {X}_{i}) / 2=\ket{+}\bra{+},  
\end{equation}
to the product initial state $\ket{+}=\otimes_i (\ket{0}_i\!+\!\ket{1}_i)/\sqrt{2}$ to boost the probability of solutions. 
Here, ${X}_i$ is the Pauli operator ${X} = \ket{1}\bra{0} + \ket{0}\bra{1}$ acting on the $i^{\text{th}}$ qubit.
This results in the state
\begin{equation}
\label{eq:qaoa_state}
\ket{\boldsymbol\beta,\boldsymbol\gamma}\equiv e^{-i\beta_p H_M} e^{-i\gamma_p H_P} \cdots e^{-i\beta_1 H_M} e^{-i\gamma_1 H_P} \ket{+},
\end{equation}
where $\boldsymbol\beta=(\beta_1,\dots,\beta_p)$ and $\boldsymbol\gamma=(\gamma_1,\dots,\gamma_p)$ are real parameters, and $p$ is the number of QAOA rounds. The Grover mixing unitary $e^{-i\beta H_M}$ for $n$ qubit problems can be decomposed with the help of $(n-3)$ ancilla qubits, as shown in Fig.~\ref{fig:qmixer} in Appendix~\ref{appendix:circuit}.

The $\boldsymbol\beta$ and $\boldsymbol\gamma$ are optimized as described in Appendix~\ref{append:search} to minimize a cost function 
\begin{equation}
\label{eq:cost}
    C \equiv \bra{\boldsymbol\beta,\boldsymbol\gamma} H_P \ket{\boldsymbol\beta,\boldsymbol\gamma}, 
\end{equation}
which can be evaluated by averaging over measurements on a quantum computer.

G-QAOA alleviates a significant obstacle to finding all solutions that confronts the usual QAOA based on the transverse-field mixer~\cite{farhi2014quantum} (X-QAOA): $H_X = \sum_{i=1}^n {X}_i$. In particular, the X-QAOA fails to sample all solutions equally, often exhibiting exponentially small weights on some solutions~\cite{Zhu2022Multi-roundComputer,matsuda2009ground}. 

\medskip
\noindent {\bf Single-angle-pair G-QAOA.} 
Optimizing many parameters in variational quantum algorithms poses difficulties to classical optimizers~\cite{bittel2021training, shaydulin2019multistart,mcclean2018barren}. 
Hence, this paper also considers a protocol in which the same pair of $(\beta,\gamma)$ angle values is used across all rounds of the G-QAOA. Accordingly, the protocol in Eq.~\eqref{eq:qaoa_state} becomes
\begin{equation}
\label{eq:const-angle}
\ket{\beta,\gamma,p}\equiv \prod_{j=1}^p e^{-i\beta H_M} e^{-i\gamma H_P} \ket{+}
\end{equation}
for $p$-round G-QAOA, reducing the number of variational parameters from $2p$ to 2. This simplification significantly alleviates the classical optimization demands inherent to the QAOA protocol.
Tests (see Sec.~\ref{sec:result}) across different problem sizes comparing the single-angle-pair G-QAOA and full G-QAOA reveal that, despite this constraint, the single-angle-pair G-QAOA exhibits a similar quadratic speedup as the full G-QAOA.

\begin{figure*}[t]%
\centering
\includegraphics[width=0.95\textwidth]{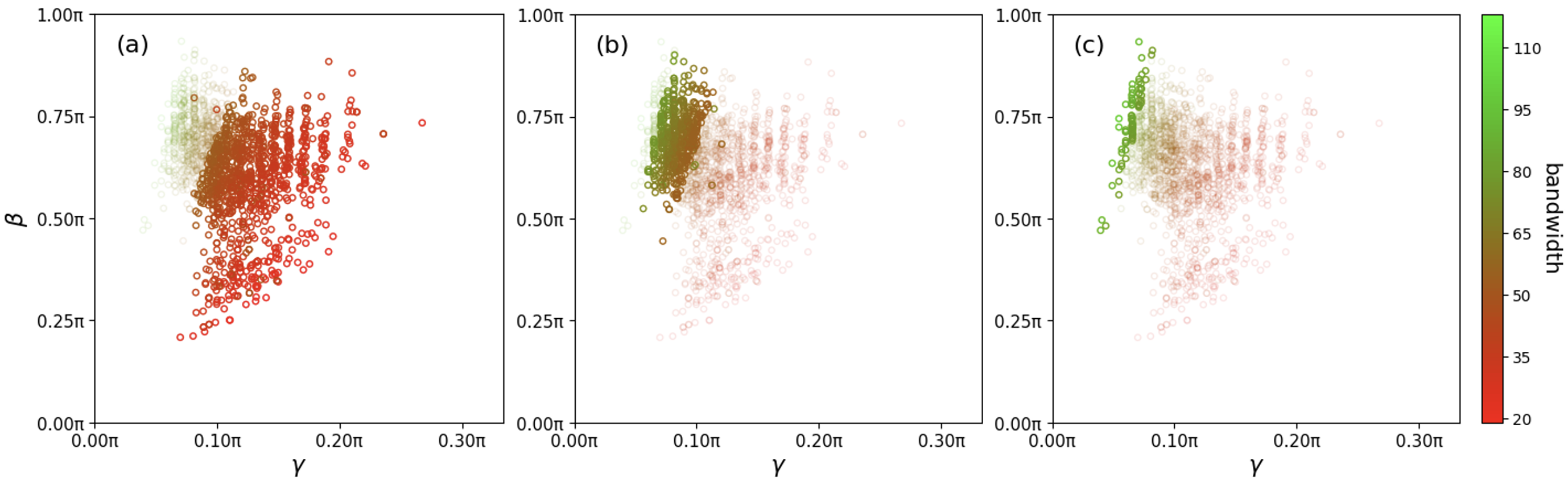}
\caption{\label{fig:clustering} \textbf{Parameter clustering}:  Distribution of angle pairs ($\beta,\gamma$) [data points in Fig.~\ref{fig: speedup}(a) for 10 to 26 qubits] in the single-angle-pair protocol, color-marked with the ``bandwidth" $m$, a product of the problem size and clause density. Opaque data points indicate (a): $m<60$; (b): $60\leq m<90$; (c): $m\geq 90$, while data points outside these ranges are shown faded.  }
\end{figure*}

\section{Quadratic speedup and Parameter Clustering}
\label{sec:result}
This section assesses the performance of G-QAOA and its single-angle-pair variant for 3-SAT problems. We produce random instances $\phi(m,n)$ of the 3-SAT problem for two density ranges, roughly below and above the satisfiability threshold: \emph{(i)} $d\in (2,4.5)$, in which the large majority of instances are satisfiable, and \emph{(ii)} $d\in(4.6,8)$ in which most are unsatisfiable. In the former, we keep only satisfiable instances; in the latter, we keep only unsatisfiable ones to highlight potential distinctions between the two cases. The two separate density ranges chosen here help to avoid cluttering in plotting, and due to the finite-size effect, there are satisfiable instances when the density is slightly above the threshold.
In generating random instances, if an instance contains duplicate clauses, only one of the repeating clauses is retained. We consider problem sizes from $n= 10$ to 26, with at least 100 random 3-SAT instances for each $n<23$, and at least 50 random instances for each $n\ge 23$ (simulation details in Appendix~\ref{append:search}).

\subsection{Evidence of quadratic speedup} 
Fig.~\ref{fig: speedup}(a) shows the number of G-QAOA rounds required to create a quantum state with at least $S=50\%$ probability of being a solution, where each point corresponds to a sampled 3-SAT instance. While the time to find a solution with random guessing would scale as $1/P$,  where $P$ is the fraction of Boolean variable assignments that satisfy $\phi$, Fig.~\ref{fig: speedup}(a) shows that the number of G-QAOA rounds scales only as $1/\sqrt{P}$.
Appendix~\ref{append:search} shows similar results for other success probabilities, $S=25\%$ and $75\%$. The same trend and quadratic speedup hold for results obtained using the simpler single-angle-pair G-QAOA, also shown in Fig.~\ref{fig: speedup}(a). Using the single-angle-pair ansatz, Fig.~\ref{fig: speedup}(b) shows the same scaling for the as-hard or harder problem of finding the variable assignments with the maximum number of satisfied clauses (Max-SAT) when instances are not fully satisfiable. 

The quadratic speedup observed in the number of QAOA rounds is maintained after accounting for the implementation cost of the quantum circuits (number of M{\o}lmer-S{\o}rensen gates~\cite{sorensen:quantum_1999}), as shown in Fig.~\ref{fig: speedup}(c).

The same data points of Figs.~\ref{fig: speedup}(a,b)  are plotted in Fig.~\ref{fig: speedup}(d) as a function of density $d$. The results up to 26 Boolean variables indicate that, for the satisfiable 3-SAT problem, the number of G-QAOA rounds required to boost the probability of solution increases linearly as $d$ increases. The number of rounds is roughly independent of $d$ for unsatisfiable instances.

\subsection{Parameter clustering} 
Tracking the angle pairs ($\beta$,$\gamma$) associated with these 3-SAT problems, a clustering pattern is observed, as shown in Fig.~\ref{fig:clustering}. For all 3-SAT instances in Fig.~\ref{fig: speedup}(a), the distribution of the angle pairs is notably concentrated, contingent upon the size of instances and the number of clauses. 
Similar clustering behaviors are also observed for the Max-SAT problems [Fig.~\ref{fig: speedup}(b)] when applying the single-angle-pair protocol, as shown in Appendix~\ref{appendix:maxsat}.

As the number of clauses increases, the distribution of angle pairs concentrates, especially the  $\gamma$ angle. This feature suggests that searching within a smaller parameter range can reduce the classical optimization cost.

\section{G-QAOA on IonQ hardware}
\label{sec:experiment}
\begin{figure}[h]%
\centering
\includegraphics[width=0.48\textwidth]{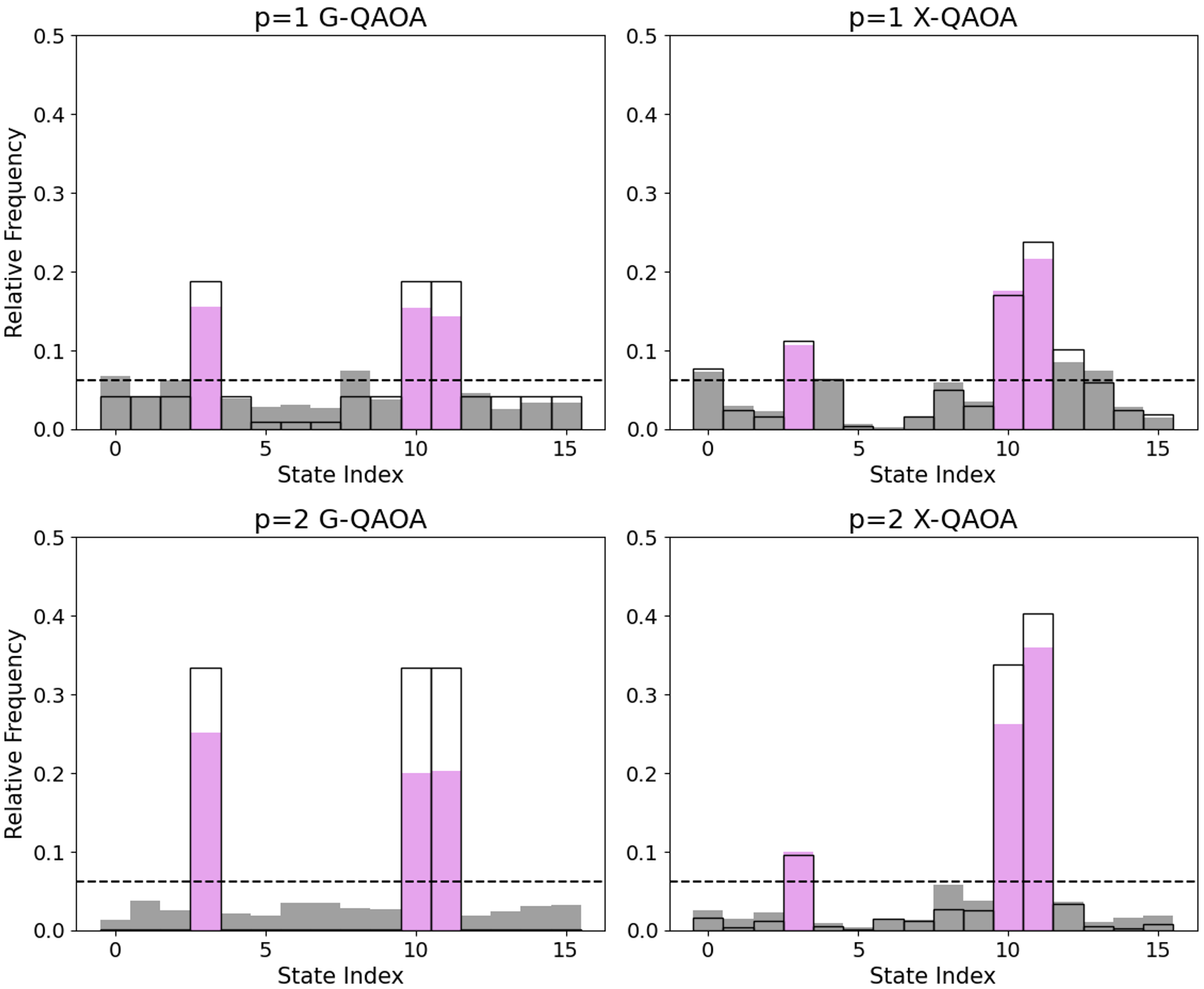}
\caption{Results from IonQ Aria machines for an $n=4$ (4 qubits + 1 ancilla) problem. Here, 2000 shots are taken for each panel. States displaying pink bins represent solutions, while the rest give non-solutions. The dashed line indicates the probability of finding a solution by random guessing. Open bins give noiseless simulated results for reference. The state index is the decimal number corresponding to the binary string for the quantum state.}\label{fig:aria_1}
\end{figure}

\begin{figure}[h]%
\centering
\includegraphics[width=0.48\textwidth]{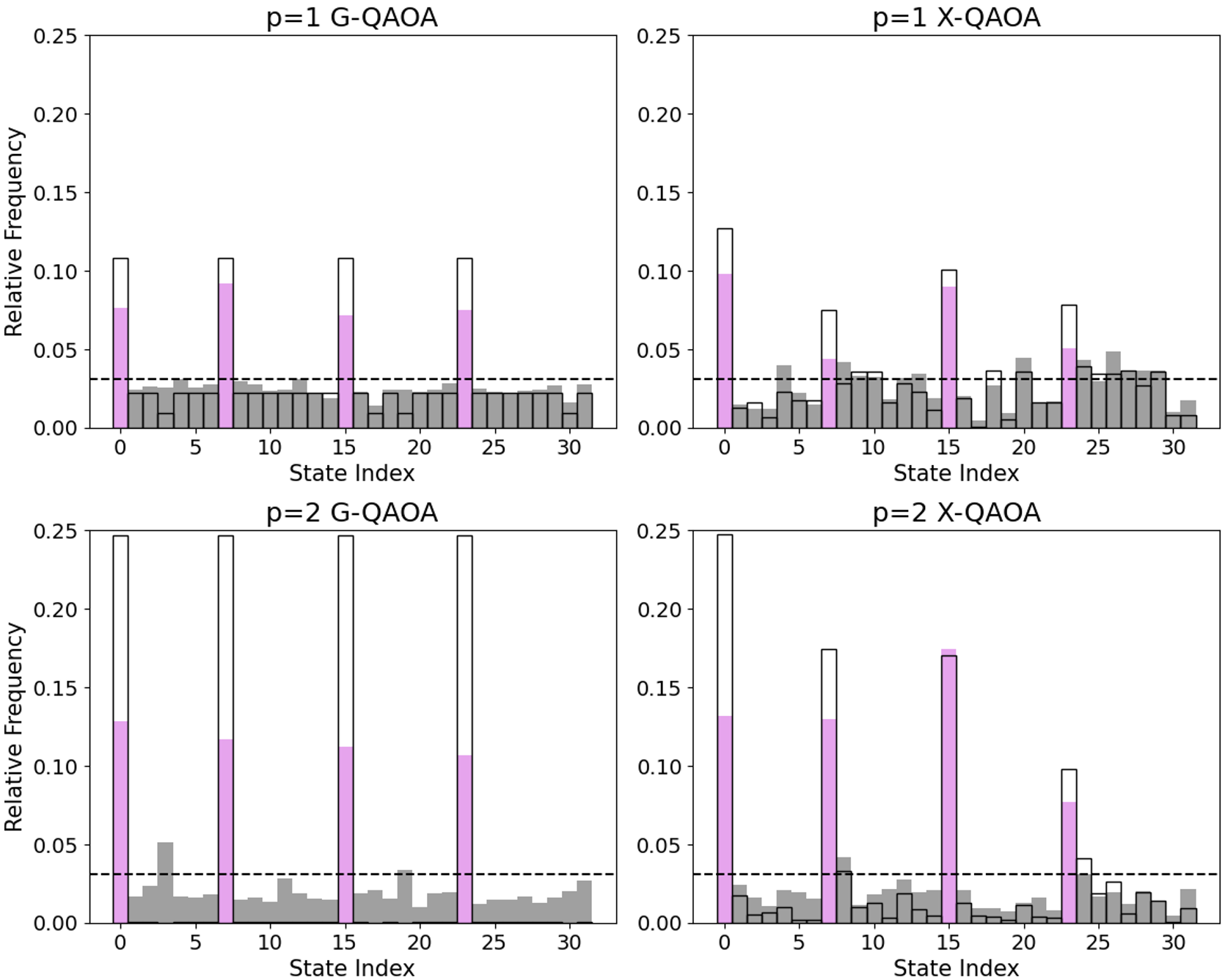}
\caption{Results from IonQ Aria machines for an $n=5$ (5 qubits +2 ancillas) problem. Here, 4,000 shots are taken for each panel. States displaying pink bins represent solutions, while the rest give non-solutions.  The dashed line indicates the probability of finding a solution by random guessing. Open bins give noiseless simulated results for reference. The state index is the decimal number corresponding to the binary string for the quantum state.}\label{fig:aria}
\end{figure}

To benchmark the feasibility of G-QAOA on NISQ devices, we executed sample 3-SAT problems (see \footnote{4-qubit example: \{[-0, +2, +3], [+0, +2, -3], [-1, +2, -3], [-1, -2, -3], [-1, -2, +3], [+1, +2, -3], [+0, +2, +3], [-0, +1, -3]\}; 5-qubit example: \{[+1, -2, -3], [-1, -3, +4], [+0, -2, -4], [-0, -2, +3], [+2, +3, +4], [-0, +1, +2], [+0, -2, 4], [-1, +2, -4]\}.
This notation lists the set of conjunctive clauses, where the notation for each clause is that minus signs imply the negation of a Boolean variable, and the number indexes the Boolean variable. For example $[-0, +2, +3]=(\bar{x}_0\vee x_2 \vee x_3)$. })
using IonQ’s Aria machine. The demonstrations were executed at optimal G-QAOA parameters that were pre-optimized classically.
The 4-qubit (Fig.~\ref{fig:aria_1}) and the 5-qubit (Fig.~\ref{fig:aria}) examples show that the G-QAOA enhances solution probability over random guessing, with each solution having a comparable relative frequency. In contrast, the X-QAOA enhances solution probabilities with relative frequencies varying significantly between solutions. 

The experimental results qualitatively agree with the noiseless numerical simulations. For larger problems ($n>5$) and rounds of G-QAOA ($p\geq3$), the results depart significantly from the noiseless simulation, especially for G-QAOA which requires considerably more entangling gates (M{\o}lmer-S{\o}rensen) than its X-QAOA counterpart at the tested densities. However, the relative difference in gate cost will decrease at higher densities.
Specifically, given $n$ Boolean variables and the number of clauses $m$, X-QAOA requires $6 m$ entangling gates per round. In contrast, G-QAOA requires $6m+5(2n-5)$ per round, where the additional $5(2n-5)$ entangling gates come from the decomposition of the Toffoli gates needed to implement the Grover mixer in Eq. \eqref{eq:Grover_Mx}. Additionally, G-QAOA requires $n-3$ ancillas to implement the Grover mixer (see Appendix~\ref{appendix:circuit} for details).  

Simulations and demonstration on Aria highlight two aspects of G-QAOA's efficacy, whose relative importance differs in different applications: the total probability of solutions and how uniformly the solutions are sampled. We utilize two metrics to assess the latter: $i)$ the number of shots needed to observe all solutions~\cite{Zhu2022Multi-roundComputer}; and $ii)$ the number of draws from the experimental solution distribution required to reject (by $\chi^2$-tests) the null hypothesis of uniform solution sampling,  taking the median from  multiple rejection trials,  a metric introduced in Ref.~\cite{golden2022fair}. 
Results from both metrics are 
shown in Table~\ref{tab}.

\begin{table}[h]
\begin{tabular}{@{}ccccc@{}}
\toprule
   \textbf{4-qubit}   & \begin{tabular}{@{}c@{}}Solution $\%$\\ (noiseless)\end{tabular} & \begin{tabular}{@{}c@{}}Solution $\%$\\ (IonQ Aria)\end{tabular}  & \begin{tabular}{@{}c@{}}Shots-to-\\all-solutions\end{tabular}
  &  \begin{tabular}{@{}c@{}}Shots-to-\\reject-fairness\end{tabular}\\
p=1 X  & 52.0 & 49.70   & 12.53(3)  & 64  \\
p=2 X  & 83.7 & \textbf{72.20}    & 11.42(3)  & 24 \\
p=1 G  & 56.1 & 45.35    &12.16(2)   &\textbf{3780}\\
p=2 G  & \textbf{99.9} & 65.30   &\textbf{8.56(1)}  & 426 \\
  \textbf{5-qubit} & \phantom{\begin{tabular}{@{}c@{}}Solution $\%$\\ (noiseless)\end{tabular}} &   &  & \\ 

p=1 X  & 38.1 & 28.18   & 35.51(7)  & 52  \\
p=2 X  & 68.9 & \textbf{51.25}   & 18.61(4)  & 78 \\
p=1 G  & 43.1 & 31.48   &26.79(5)  & 628 \\
p=2 G  & \textbf{98.7} & 46.40  & \textbf{18.09(3)}  & \textbf{1278} \\
\botrule
\end{tabular}
\caption{\label{tab} \textbf{Performance of X- and G-QAOA on IonQ Aria machines with 4 and 5 qubits.} Results are summarized from Fig.~\ref{fig:aria_1} and Fig.~\ref{fig:aria}. Shots-to-all-solutions: values are measured from 100,000 independent runs, and the indicated uncertainty is the 95\% confidence interval.
Shots-to-reject-fairness: values are taken from 10,000 trials. Bold values indicate the best-performing algorithm in each column. The metrics in each column are defined in the main text. 
}
\end{table}

While noiseless G-QAOA outperforms its X-QAOA counterpart in all metrics, results from Aria show that the X-QAOA solution probability is often larger than the G-QAOA's since the G-QAOA requires greater circuit depth per round.  Despite this, the G-QAOA outperforms the X-QAOA in both shots-to-all-solutions and shots-to-reject-fairness metrics. 
This shows that G-QAOA's fair-sampling advantage is evident even on noisy devices.

With both X-QAOA and G-QAOA, the experimental data also shows an enhancement in solution probability from $p=1$ to $p=2$. While such improvement for X-QAOA is common~\cite{harrigan2021quantum,Zhu2022Multi-roundComputer,shaydulin2023qaoa,moses:race-track_2023}, seeing the round-to-round improvement for G-QAOA had been observed only in Ref.~\cite{Zhu2022Multi-roundComputer} for the edge cover problem, which is less computationally challenging.

\section{Conclusion}
\label{sec:conclusion}
Our research reports numerical evidence of the quadratic speedup of G-QAOA over random guessing for All-SAT and All-Max-SAT problems for thousands of tested 3-SAT instances. The sampled instances cover problems with 10 to 26 Boolean variables across various densities.  
Compared to the Grover algorithm, which also gives a quadratic speedup, the G-QAOA for 3-SAT problems requires less quantum resources in both circuit depth and auxiliary qubits. It is also more flexible and can be directly applied to Max-SAT problems. 
We do note that in contrast to random sampling, leading classical algorithms, as detailed in Appendix~\ref{appendix:classical}, may scale better than G-QAOA for typical instances; however, for hard instances in which the classical algorithms are likely to saturate the upper bound~\cite{DECHTER199941},
we are optimistic that G-QAOA will maintain a quadratic advantage since it does not rely in any obvious way on the structure of the problem.
Furthermore, we execute X-QAOA and G-QAOA for small $4-$ and $5-$variable instances on the IonQ Aria quantum computing platform, showing that the hardware maintains the uniform solution sampling property for G-QAOA relative to X-QAOA as quantified by reduced shots-to-all-solutions and other fairness metrics. Despite the longer circuit depth, we also note that the G-QAOA results obtained from demonstrations on the IonQ Aria machine improve from levels $p=1$ to $p=2$.

We also observe that a single-angle-pair constraint provides a simplified G-QAOA protocol that requires optimizing only two parameters while maintaining the quadratic speedup seen in the full G-QAOA.  The faster optimization of the single-angle-pair approach allows us to obtain $26$-qubit numerical results with comparable computational time required to obtain $20$-qubit results for the full G-QAOA on classical simulators. 

We also find clustering of the two parameters in the single-angle-pair G-QAOA, even for a large number of rounds $p\gtrsim 1000$. 
While parameter clustering has previously been seen in different QAOA variants, these focused on only few-round results~\cite{streif2020training,akshay2021parameter,moussa2022unsupervised}. 
Therefore, the findings in this work are promising for future large-scale applications of QAOA, and can significantly reduce the classical overhead on parameter training. For example, it possibly could circumvent the notorious barren plateaus challenge in variational searches~\cite{mcclean2018barren,wang2021noise}.
While quadratic speedups are unlikely to be sufficient for practical applications in the foreseeable future \cite{babbush2021,beverland2022assessing}, our results are likely to give insight into designing quantum algorithms more generally.

This research lays the groundwork for progress in several directions. These include understanding the observed efficacy of the single-angle-pair and clustering in the G-QAOA protocol, using parameter clustering to accelerate the optimization of the G-QAOA parameters further, applying G-QAOA to other problems in science and engineering, and exploring avenues for achieving more significant speedup. Given the fundamental role of SAT problems, it encourages future work to apply this single-angle-pair protocol to other combinatorial optimization problems. 
We believe a fruitful area suggested by these results is to find simple, possibly analytical, ways of determining the G-QAOA angles from the problem structure, for example, statistical characteristics of its density of states. See Ref.~\cite{headley2023problem,bridi2024analytical} for related considerations. Finally, other questions are to what extent the run time may improve when we soften the single-angle-pair constraint differently from the na\"ive relaxation, or when use a warm-start~\cite{egger2021warm,tate2023warm}.

\begin{acknowledgments}
The authors thank Dr. Nai-Hui Chia for providing valuable insights. 
G.P., K.R.A.H. and Z.Z. acknowledge support from the Office of Naval Research (N00014-20-1-2695 and N00014-23-12665).
K.R.A.H. and Z.Z. also acknowledge support from the National Science Foundation (PHY-1848304 and CMMI-2037545) and the Department of Energy (DE-SC0024301). 
G.P. also acknowledges support from the National Science Foundation (NSF CAREER award No. PHY-2144910), the Army Research Office (Grant No. W911NF22C0012), the Welch Foundation Grant No. C-2154, and the Office of Naval Research Young Investigator Program (N00014-22-1-2282). G.P. acknowledges that this material is based upon work supported by the U.S. Department of Energy, Office of Science, Office of Nuclear Physics under the Early Career Award No. DE-SC0023806.
A. K. and D. Q. acknowledge support by NSF FET:Small No. 1907936, NSF CMMI No. 2037545, NSF CAREER award No. 2145629, Welch Foundation Grant \#A22-0307, Microsoft Research Award, Amazon Research Award, and a Rice InterDisciplinary Excellence Award (IDEA). L.D.O. and R.P. acknowledge support by NSF award CMMI-2037545.

\end{acknowledgments}

\appendix

\section{Comparing with Grover algorithm}
\label{append:ga}
For 3-SAT problems, the Grover algorithm is more costly than the G-QAOA in ancilla number and circuit depth. Building up the Grover oracle for $\phi(m,n)$ takes $2m$ triple-controlled gates for the \texttt{OR} logic and two $m$-controlled gates for the \texttt{AND} logic, which can roughly be mapped to $10m$ Toffoli gates or $50m$ MS gates.  One ancilla is needed for each clause as the register for \texttt{OR} logic, and $m$ ancillas are needed to decompose the $m$-controlled gate, which gives a total of $2m$ ancillas. Thus, around the critical density $m\sim 4.26$, the QAOA problem unitary reduces the circuit depth and scalability by an order of magnitude. Meanwhile, the Grover algorithm requires $\frac{\pi}{8\sqrt{P}}$ iterations to boost the solution probability to $\frac{1}{2}$, where the constant is close to what we observe for G-QAOA in Fig.~\ref{fig: speedup}(a) and (b), \textit{i.e.}, at large $\frac{1}{P}$ it takes roughly $\frac{1}{3\sqrt{P}}$ rounds of G-QAOA to boost the solution probability to $\frac{1}{2}$. 

The benefits of G-QAOA become more pronounced at higher densities, even though 3-SAT problems are less likely to possess satisfying solutions in such regimes. Nonetheless, identifying states that maximize the number of satisfied clauses (Max-SAT) is also a critical and hard problem. As the density $m$ increases and solutions become sparse, the Grover algorithm may require relatively more resources due to the absence of solutions with large $m$. In these high-density regions, G-QAOA may most significantly outperform the traditional Grover algorithm in gate counts.

\section{The Boolean satisfiability problem}
\label{appendix:sat}
A Boolean formula $\phi\equiv\phi(m,n)$ over a set $X$ of $n$ Boolean variables is in CNF when it is of the form
\begin{equation}
    \phi = C_1 \wedge C_2 \wedge \dots \wedge C_m,
\end{equation}
where  $\wedge$ denotes logical conjunction (\texttt{AND}), and every clause, $C_i$, $1\leq i\leq m $, is of the form
\begin{equation}
    C_i = l_{i_1} \vee l_{i_2} \vee \dots  \vee l_{i_{k_i}},
\end{equation}
where  $\vee$ represents  logical disjunction (\texttt{OR}), $k_i$ is the length of clause $C_i$, and every literal $l_{i_j}$ is either equal to a variable $x\in X$ or its negation  $\bar{x}$. 
We use $s_{i_j}$ to denote the sign of literal $l_{i_j}$, \textit{i.e.},  $s_{i_j}=1$ for $x_{i_j}$ and $s_{i_j}=-1$ for $\bar{x}_{i_j}$.

An assignment of truth values to variables (or ``configuration"), is a map   $\tau\colon X\to \{0, 1\}$. 
For $x\in X$, we say that an assignment $\tau$ satisfies the literal $x$ when $\tau(x)=1$, and satisfies the literal $\bar{x}$ when $\tau(x)=0$. 
An assignment $\tau$ satisfies clause $C_i$ if $\tau$ satisfies at least one literal $l_{i,j}$, $1\leq j \leq k_i$, and $\tau$ satisfies formula $\phi$ when $\tau$ satisfies every clause $C_i$, $1\leq i\leq m$. 
For convenience, we write $\phi(\tau)=1$ when $\tau$ satisfies $\phi$ and write $\phi(\tau)=0$, otherwise.

The Boolean satisfiability (SAT) problem is deciding whether an arbitrary Boolean formula $\phi$ expressed in CNF has at least one satisfying assignment. 
A \textit{solution} is a satisfying assignment of the formula, and we use $R_\phi=\{\tau: \phi(\tau)=1\}$ to denote the set of solutions of $\phi$. 
The satisfying probability, i.e., the likelihood that a random assignment is a solution, is $P=\vert R_{\phi} \vert/ 2 ^n$. 

Several crucial combinatorial optimization problems are variants of the Boolean satisfiability problem. If all the clauses have $k$ literals, deciding whether there is a satisfying assignment is the $k$-SAT problem\cite{cook2023complexity,karp2010reducibility}. Finding all the satisfying assignments is All-SAT~\cite{yu2014all}. Finding an assignment with the fewest unsatisfied clauses (which could be non-zero) is known as Max-SAT~\cite{asano2002improved,nannicini2019performance}. Finding all assignments that have the fewest unsatisfied clauses is known as Max-All-SAT. All are important in theoretical computer science and have practical applications in science and engineering~\cite{duenas2017counting}.

\section{Optimizing QAOA angles}
\label{append:search}

\begin{figure}[t]
\centering
\includegraphics[width=0.75\linewidth]{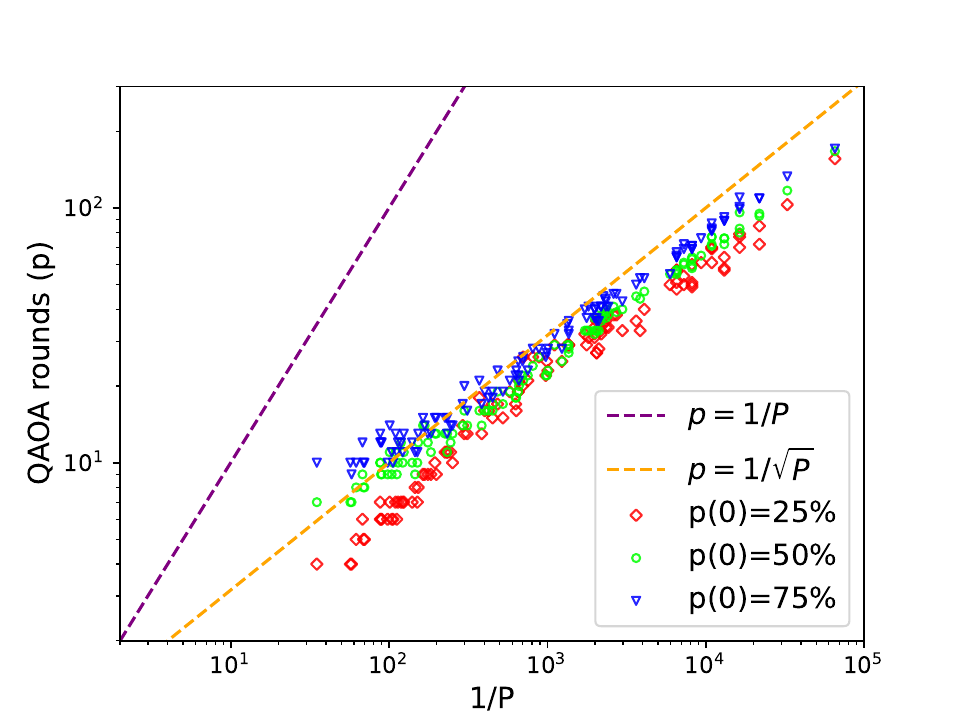}
\caption{\label{fig:convergence}  Same 16-variable instances considered with G-QAOA in the main text for different target solution ratios for the final state, 25\%, 50\%, and 75\%. }
\end{figure}

\noindent\textbf{Optimization.} To minimize the 3-SAT cost function, Eq.~\eqref{eq:cost}, for single-angle-pair G-QAOA, we first evaluate the cost function on a uniform grid in $\beta$ and $\gamma$ with grid spacing $\pi/180$ in each direction for problem size no larger than 20, while for larger problems we sample 4000 angles per instance in the $(\beta,\gamma)$-plane. We then locally optimize the $(\beta,\gamma)$ using these values as initial parameters for \textit{L-BFGS-B} as implemented in the \textit{scipy.optimize} package. We use a stopping criterion of $10^{-6}$ relative changes of the cost function between consecutive iterations.  For the G-QAOA results where the angles $(\beta_p,\gamma_p)$ are allowed to depend on the round, we use the same initial seeds but allow the  \textit{L-BFGS-B} local relaxation to optimize all the angles. 

In generating random instances, for each independent $n$-variable sample, we uniformly select a value $d$ within the density range, and generate $nd$ random clauses according to the density. Should 
$nd$ not result in an integer, appropriate rounding is applied. During the classical simulation, we leverage the G-QAOA's invariance under the relabeling of states within the same energy manifold. This simplification enables us to do the exact simulation into two steps. The first step is a classical pre-computation which requires counting the states in each distinct energy manifold, which takes $O(2^n)$ computational time for $n$ Boolean variables. The second step involves exact classical simulation of the quantum circuit, but this simulation is not in the full Hilbert space of dimension $2^n$, and instead is done in an effective space whose dimension is the number of distinct energy manifolds $\sim O(m)$~\cite{sundar2019quantum}. We are therefore able to simulate the G-QAOA for up to 26 qubits, with one of the main bottlenecks being the classical pre-computation.

\noindent\textbf{Convergence.} 
The choice of 50\% solution probability at the end of the QAOA  does not affect the quadratic trend in Fig.~\ref{fig: speedup}. For example, the 16-variable instances of 3-SAT are tested in Fig.~\ref{fig:convergence} with two different ending solution ratios below and above 50\%. As shown in the graph, different choices of ending ratios for the target states do not impact on the general scaling.

\begin{figure}[t]
\centering
\includegraphics[width=0.99\linewidth]{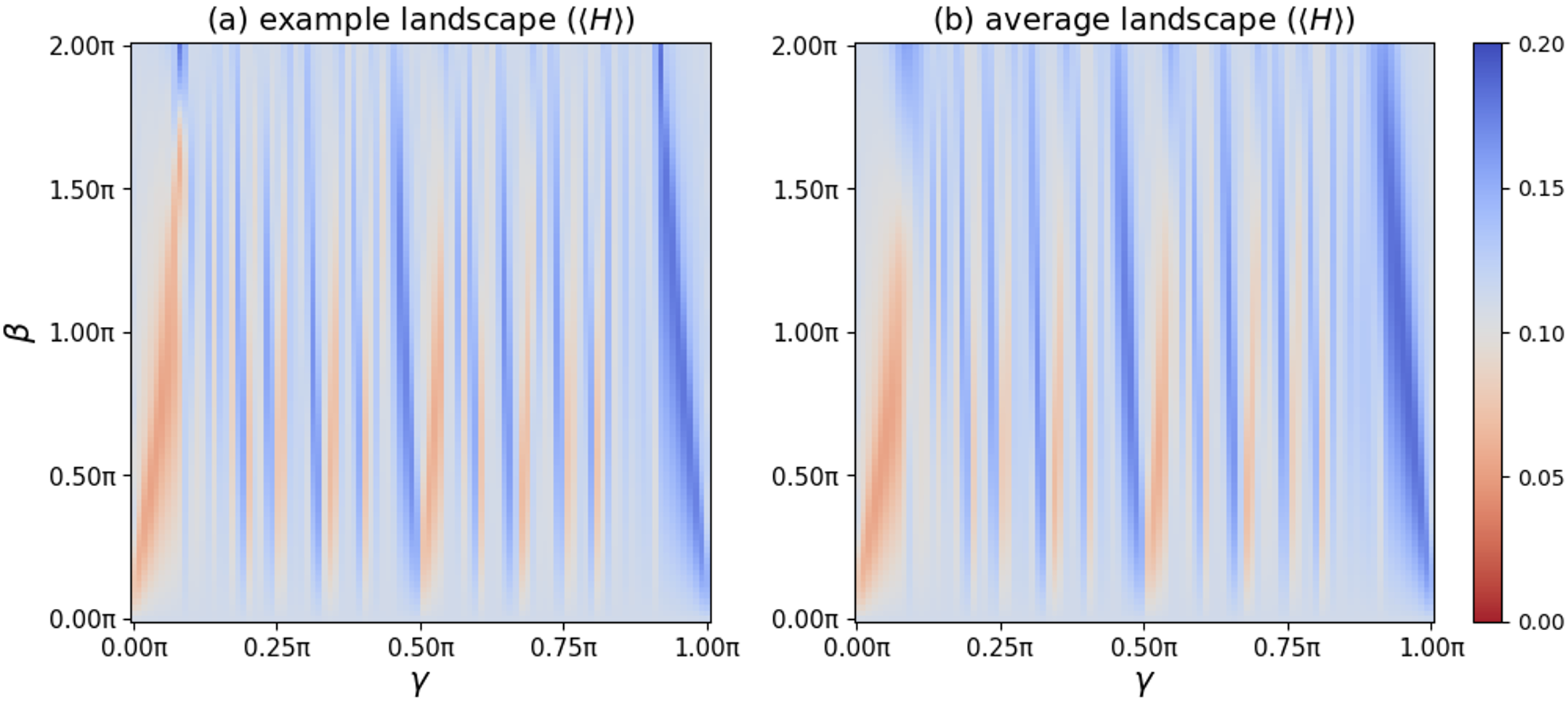}
\caption{\label{fig:landscape} (a) The single-angle-pair G-QAOA landscape [energy $\langle H\rangle$ according to~\eqref{eq:cost}] of a typical 10-qubit instance at $p$=10. (b) Average landscape of all 10-qubit instances at $p$=10.}
\end{figure}

\noindent\textbf{Landscape.} 
To illustrate the optimization landscape of the single-angle-pair G-QAOA for 3-SAT,  Fig.~\ref{fig:landscape} shows the energy landscape on the ($\beta$,$\gamma$) surface. Fig.~\ref{fig:landscape}(a) gives an example of a typical instance. Fig.~\ref{fig:landscape}(b)  averages the energy landscape over all 108 tested instances of 10-qubit. Typical landscapes are peaked in a similar area to the average. These results indicate an area within the parameter space ($\beta$,$\gamma$) that yields the best performance, corresponding to the region of parameter clustering observed in Fig.~\ref{fig:clustering}.

\section{Parameter clustering for Max-SAT}
\label{appendix:maxsat}
\begin{figure}[h]%
\centering
\includegraphics[width=0.9\linewidth]{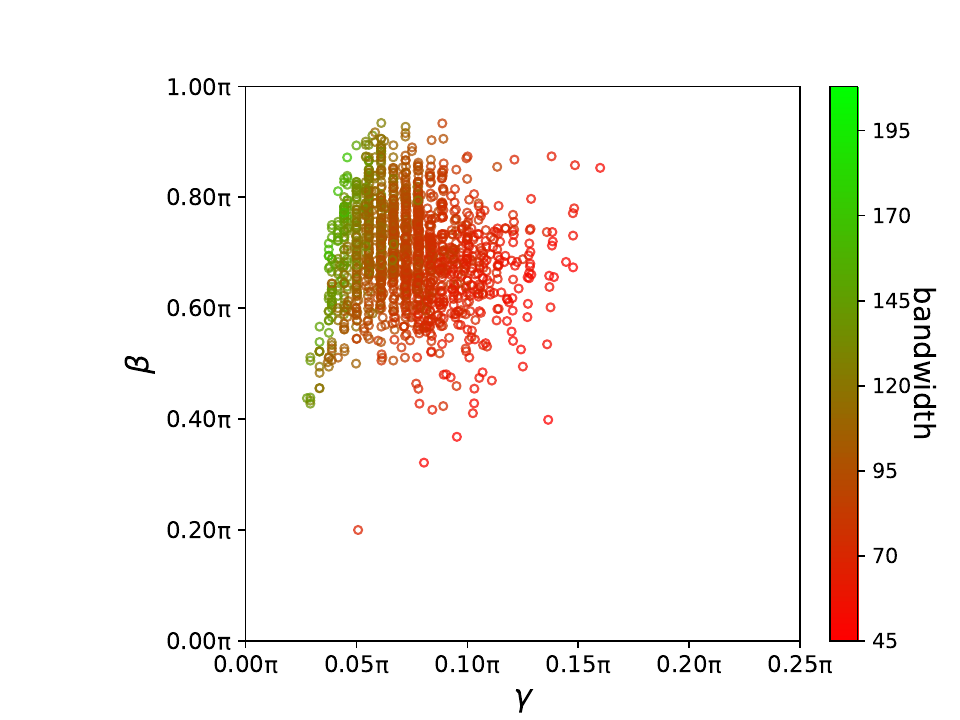}
\caption{\label{fig:maxsat-cluster}  Distribution of angle pairs ($\beta,\gamma$) [Max-SAT data points in Fig.~\ref{fig: speedup}(b) for 10 to 26 qubits] in the single-angle-pair protocol, color-marked with the number of clauses $m$, which is the product of the problem size and clause density. }
\end{figure}

The parameter clustering (the same as Fig.~\ref{fig:clustering}) is also observed for all the Max-SAT instances in Fig.~\ref{fig: speedup}(b), as shown in Fig.~\ref{fig:maxsat-cluster}.

\begin{figure*}[t]
    \centering \includegraphics[width=0.9\linewidth]{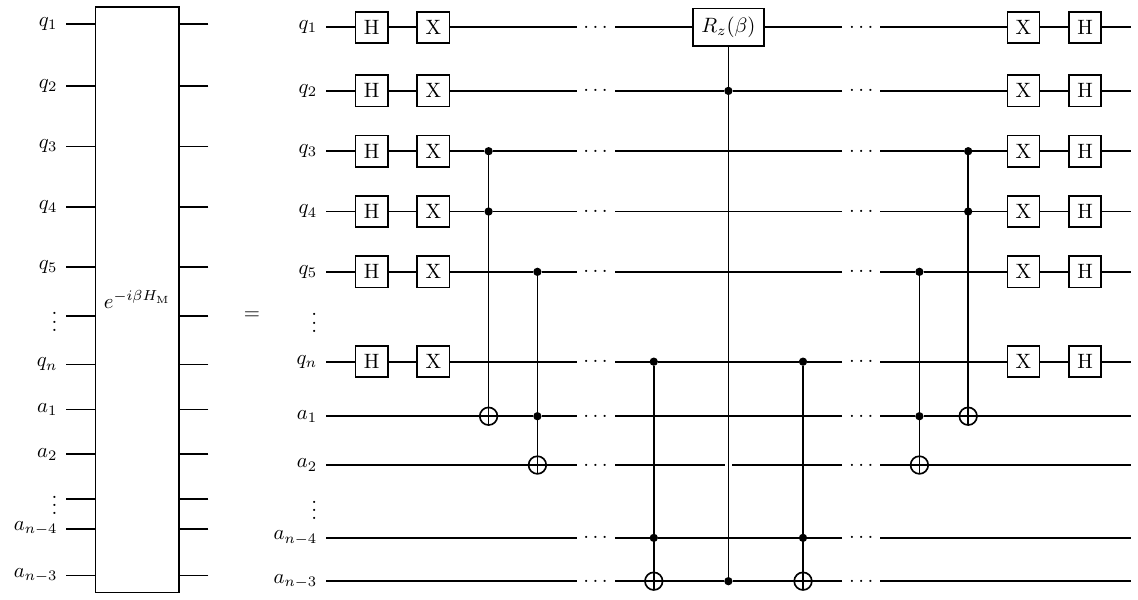}
    \caption{Grover mixer for $n$-qubit problems. For the case of $n=3$, the only entangling gate in the middle is a double-controlled phase gate, and no ancilla is needed.
    }
    \label{fig:qmixer}
\end{figure*}

\section{Circuit compilation}
\label{appendix:circuit}
The $n$-qubit Grover mixer~\cite{sundar2019quantum} can be decomposed into multiple double-controlled gates, as shown in Fig.~\ref{fig:qmixer}. Further mapping double-controlled gates onto native gates of trapped-ion machines, the M{\o}lmer-S{\o}rensen (MS) gates, is done as described in Ref.~\cite{maslov2017basic}.

\begin{figure*}
    \centering   \includegraphics[width=0.65\linewidth]{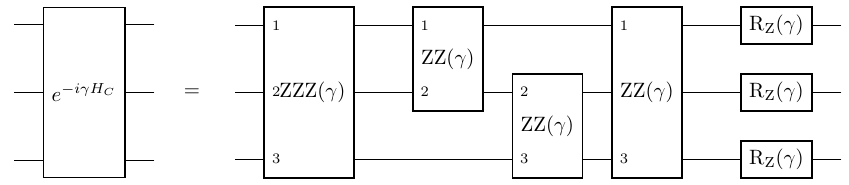}
    \caption{A sample problem unitary for a 3-SAT clause. The signs of gates depend on the clause to encode. Note the $\mathrm{ZZ}$ gate here follows the convention of \textit{braket}, which differs from that in the trapped-ion community by a factor of 2. }
    \label{fig:prob_U}
\end{figure*}

An illustrative problem unitary for clause $[+1,+2,+3]$ is encoded as shown in Fig.~\ref{fig:prob_U}.

When running circuits on the Aria machines,  to reduce the impact of possible local noises, we reassigned the order of encoding qubits every 500 shots during the demonstration on Aria.

\begin{figure*}
\centering
\includegraphics[width=0.99\textwidth]{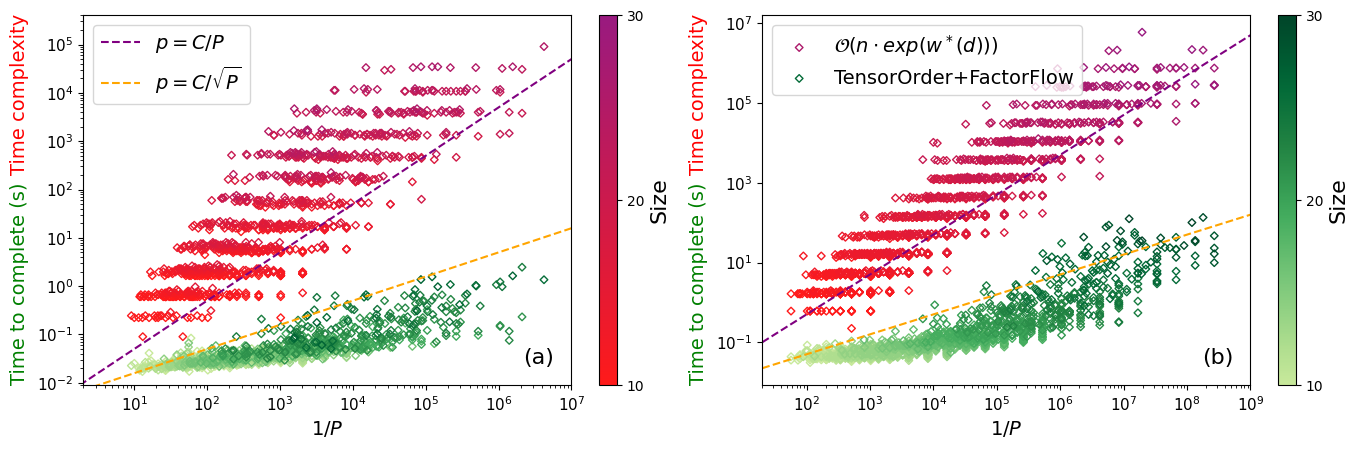}
\caption{\textbf{Classical scaling}: Running time for \textit{TensorOrder} using \textit{FactorFlow} in seconds and upper bound on time complexity of the \textit{Elim-Count} algorithm given the treewidth of each CNF for the instances of Fig.~\ref{fig: speedup} in \textbf{(a)}, and for randomly generated instances up to 30 bits with density $d \approx 4.26$ in \textbf{(b)}. The red data points represent the time complexity of the \textit{Elim-Count} algorithm given the treewidth calculated by \textit{FactorFlow} to enumerate solutions, while the green data points represent the total time to count 3-SAT solutions using \textit{FactorFlow} in seconds, but without enumerating them.}\label{fig:classical_results}
\end{figure*}

\section{Comparison with classical algorithms}
\label{appendix:classical}
Classical approaches to 3-SAT problems such as branch and bound~\cite{liefficient2010} algorithms, tensor networks~\cite{gray_hyperoptimized_2021,dudek_parallel_2021,tamaki_positive-instance_2018,hamann_graph_2017, hicks2002branchwidth}, and weighted model counting~\cite{nagy_ising_2022,dudek_parallel_2021}, provide various important comparisons for the G-QAOA results. 
Among these, we analyze the performance of a bucket elimination algorithm \textit{Elim-Count}~\cite{Dechter2002GeneratingRS} which finds all solutions. An upper bound on the time complexity of the \textit{Elim-Count} algorithm is $\mathcal{O}(n\cdot \text{exp}(w^*(d)))$, with $n$ nodes and $w^*(d)$ induced treewidth, related to the treewidth $w$ by $w^*(d)=w-1$. \textit{Elim-Count} shares the same exponential time and space complexity on the induced width of the problem graph as other bucket-elimination algorithms~\cite{DECHTER199941}. It can also aid in sampling instances, a task that is at least as hard as counting. Sampling through \textit{Elim-Count} can be performed in time growing linearly with the number of possible assignments by accessing the underlying graph constructed during the bucket elimination process. 

We also compare runtimes of our algorithm to results obtained from \textit{TensorOrder}~\cite{dudek_parallel_2021}, a weighted model counting tool that operates by contracting  tensor networks. \textit{TensorOrder} can use  different solvers to count the number of assignments that satisfy a given CNF. We use \textit{FlowCutter} as the classical solver as it provides the best performance. \textit{FlowCutter} optimizes cut size and undergoes network planning, tree construction, and tree contraction phases where the optimal contraction is sought with a time complexity of $\mathcal{O}(c|E|)$, with $c$ the optimal cut size and $|E|$ the number of edges. It also enables us to calculate the treewidth for determining the upper bound of the scaling of \textit{Elim-Count}.

Fig.~\ref{fig:classical_results}(a) illustrates the scaling of  \textit{Elim-Count} and \textit{TensorOrder} on the 3-SAT instances used in Fig.~\ref{fig: speedup}(a), while Fig.~\ref{fig:classical_results}(b) provides additional hard instances up to 30 variables around the critical density $d=4.26$. We use \textit{FlowCutter} to calculate the run time and treewidth of each instance, and supply the calculated treewidth using the upper bound on time complexity scaling of \textit{Elim-Count}. By comparing the scaling of these classical algorithms with G-QAOA, both in complexity and runtime, we observe that \textit{TensorOrder} gives better scaling than G-QAOA on typical random instances when focusing on counting but not enumerating solutions as \textit{Elim-Count} does. However, for instances around the critical density in Fig.~\ref{fig:classical_results}(b), although at small $1/P$ the performance of \textit{TensorOrder} is better than G-QAOA, the performance at large $1/P$ is comparable to G-QAOA, although \textit{TensorOrder} does not sample solutions as G-QAOA ends up doing in its implementation.

\bibliography{apssamp}

\begin{thebibliography}{91}%
\makeatletter
\providecommand \@ifxundefined [1]{%
 \@ifx{#1\undefined}
}%
\providecommand \@ifnum [1]{%
 \ifnum #1\expandafter \@firstoftwo
 \else \expandafter \@secondoftwo
 \fi
}%
\providecommand \@ifx [1]{%
 \ifx #1\expandafter \@firstoftwo
 \else \expandafter \@secondoftwo
 \fi
}%
\providecommand \natexlab [1]{#1}%
\providecommand \enquote  [1]{``#1''}%
\providecommand \bibnamefont  [1]{#1}%
\providecommand \bibfnamefont [1]{#1}%
\providecommand \citenamefont [1]{#1}%
\providecommand \href@noop [0]{\@secondoftwo}%
\providecommand \href [0]{\begingroup \@sanitize@url \@href}%
\providecommand \@href[1]{\@@startlink{#1}\@@href}%
\providecommand \@@href[1]{\endgroup#1\@@endlink}%
\providecommand \@sanitize@url [0]{\catcode `\\12\catcode `\$12\catcode `\&12\catcode `\#12\catcode `\^12\catcode `\_12\catcode `\%12\relax}%
\providecommand \@@startlink[1]{}%
\providecommand \@@endlink[0]{}%
\providecommand \url  [0]{\begingroup\@sanitize@url \@url }%
\providecommand \@url [1]{\endgroup\@href {#1}{\urlprefix }}%
\providecommand \urlprefix  [0]{URL }%
\providecommand \Eprint [0]{\href }%
\providecommand \doibase [0]{https://doi.org/}%
\providecommand \selectlanguage [0]{\@gobble}%
\providecommand \bibinfo  [0]{\@secondoftwo}%
\providecommand \bibfield  [0]{\@secondoftwo}%
\providecommand \translation [1]{[#1]}%
\providecommand \BibitemOpen [0]{}%
\providecommand \bibitemStop [0]{}%
\providecommand \bibitemNoStop [0]{.\EOS\space}%
\providecommand \EOS [0]{\spacefactor3000\relax}%
\providecommand \BibitemShut  [1]{\csname bibitem#1\endcsname}%
\let\auto@bib@innerbib\@empty
\bibitem [{\citenamefont {Brassard}\ \emph {et~al.}(2002)\citenamefont {Brassard}, \citenamefont {Hoyer}, \citenamefont {Mosca},\ and\ \citenamefont {Tapp}}]{brassard2002quantum}%
  \BibitemOpen
  \bibfield  {author} {\bibinfo {author} {\bibfnamefont {G.}~\bibnamefont {Brassard}}, \bibinfo {author} {\bibfnamefont {P.}~\bibnamefont {Hoyer}}, \bibinfo {author} {\bibfnamefont {M.}~\bibnamefont {Mosca}},\ and\ \bibinfo {author} {\bibfnamefont {A.}~\bibnamefont {Tapp}},\ }\bibfield  {title} {\bibinfo {title} {Quantum amplitude amplification and estimation},\ }\href@noop {} {\bibfield  {journal} {\bibinfo  {journal} {Contemp. Math.}\ }\textbf {\bibinfo {volume} {305}},\ \bibinfo {pages} {53} (\bibinfo {year} {2002})}\BibitemShut {NoStop}%
\bibitem [{\citenamefont {Kadowaki}\ and\ \citenamefont {Nishimori}(1998)}]{kadowaki1998quantum}%
  \BibitemOpen
  \bibfield  {author} {\bibinfo {author} {\bibfnamefont {T.}~\bibnamefont {Kadowaki}}\ and\ \bibinfo {author} {\bibfnamefont {H.}~\bibnamefont {Nishimori}},\ }\bibfield  {title} {\bibinfo {title} {Quantum annealing in the transverse {I}sing model},\ }\href@noop {} {\bibfield  {journal} {\bibinfo  {journal} {Phys. Rev. E}\ }\textbf {\bibinfo {volume} {58}},\ \bibinfo {pages} {5355} (\bibinfo {year} {1998})}\BibitemShut {NoStop}%
\bibitem [{\citenamefont {McClean}\ \emph {et~al.}(2016)\citenamefont {McClean}, \citenamefont {Romero}, \citenamefont {Babbush},\ and\ \citenamefont {Aspuru-Guzik}}]{mcclean2016theory}%
  \BibitemOpen
  \bibfield  {author} {\bibinfo {author} {\bibfnamefont {J.~R.}\ \bibnamefont {McClean}}, \bibinfo {author} {\bibfnamefont {J.}~\bibnamefont {Romero}}, \bibinfo {author} {\bibfnamefont {R.}~\bibnamefont {Babbush}},\ and\ \bibinfo {author} {\bibfnamefont {A.}~\bibnamefont {Aspuru-Guzik}},\ }\bibfield  {title} {\bibinfo {title} {The theory of variational hybrid quantum-classical algorithms},\ }\href@noop {} {\bibfield  {journal} {\bibinfo  {journal} {New J. Phys.}\ }\textbf {\bibinfo {volume} {18}},\ \bibinfo {pages} {023023} (\bibinfo {year} {2016})}\BibitemShut {NoStop}%
\bibitem [{\citenamefont {Hadfield}\ \emph {et~al.}(2019)\citenamefont {Hadfield}, \citenamefont {Wang}, \citenamefont {O'Gorman}, \citenamefont {Rieffel}, \citenamefont {Venturelli},\ and\ \citenamefont {Biswas}}]{Hadfield2019FromAnsatz}%
  \BibitemOpen
  \bibfield  {author} {\bibinfo {author} {\bibfnamefont {S.}~\bibnamefont {Hadfield}}, \bibinfo {author} {\bibfnamefont {Z.}~\bibnamefont {Wang}}, \bibinfo {author} {\bibfnamefont {B.}~\bibnamefont {O'Gorman}}, \bibinfo {author} {\bibfnamefont {E.~G.}\ \bibnamefont {Rieffel}}, \bibinfo {author} {\bibfnamefont {D.}~\bibnamefont {Venturelli}},\ and\ \bibinfo {author} {\bibfnamefont {R.}~\bibnamefont {Biswas}},\ }\bibfield  {title} {\bibinfo {title} {From the quantum approximate optimization algorithm to a quantum alternating operator ansatz},\ }\href@noop {} {\bibfield  {journal} {\bibinfo  {journal} {Algorithms 2019, Vol. 12, Page 34}\ }\textbf {\bibinfo {volume} {12}},\ \bibinfo {pages} {34} (\bibinfo {year} {2019})}\BibitemShut {NoStop}%
\bibitem [{\citenamefont {Farhi}\ \emph {et~al.}(2014)\citenamefont {Farhi}, \citenamefont {Goldstone},\ and\ \citenamefont {Gutmann}}]{farhi2014quantum}%
  \BibitemOpen
  \bibfield  {author} {\bibinfo {author} {\bibfnamefont {E.}~\bibnamefont {Farhi}}, \bibinfo {author} {\bibfnamefont {J.}~\bibnamefont {Goldstone}},\ and\ \bibinfo {author} {\bibfnamefont {S.}~\bibnamefont {Gutmann}},\ }\bibfield  {title} {\bibinfo {title} {A quantum approximate optimization algorithm},\ }\href@noop {} {\bibfield  {journal} {\bibinfo  {journal} {arXiv preprint arXiv:1411.4028}\ } (\bibinfo {year} {2014})}\BibitemShut {NoStop}%
\bibitem [{\citenamefont {Pagano}\ \emph {et~al.}(2020)\citenamefont {Pagano}, \citenamefont {Bapat}, \citenamefont {Becker}, \citenamefont {Collins}, \citenamefont {De}, \citenamefont {Hess}, \citenamefont {Kaplan}, \citenamefont {Kyprianidis}, \citenamefont {Tan}, \citenamefont {Baldwin}, \citenamefont {Brady}, \citenamefont {Deshpande}, \citenamefont {Liu}, \citenamefont {Jordan}, \citenamefont {Gorshkov},\ and\ \citenamefont {Monroe}}]{Pagano2020}%
  \BibitemOpen
  \bibfield  {author} {\bibinfo {author} {\bibfnamefont {G.}~\bibnamefont {Pagano}}, \bibinfo {author} {\bibfnamefont {A.}~\bibnamefont {Bapat}}, \bibinfo {author} {\bibfnamefont {P.}~\bibnamefont {Becker}}, \bibinfo {author} {\bibfnamefont {K.~S.}\ \bibnamefont {Collins}}, \bibinfo {author} {\bibfnamefont {A.}~\bibnamefont {De}}, \bibinfo {author} {\bibfnamefont {P.~W.}\ \bibnamefont {Hess}}, \bibinfo {author} {\bibfnamefont {H.~B.}\ \bibnamefont {Kaplan}}, \bibinfo {author} {\bibfnamefont {A.}~\bibnamefont {Kyprianidis}}, \bibinfo {author} {\bibfnamefont {W.~L.}\ \bibnamefont {Tan}}, \bibinfo {author} {\bibfnamefont {C.}~\bibnamefont {Baldwin}}, \bibinfo {author} {\bibfnamefont {L.~T.}\ \bibnamefont {Brady}}, \bibinfo {author} {\bibfnamefont {A.}~\bibnamefont {Deshpande}}, \bibinfo {author} {\bibfnamefont {F.}~\bibnamefont {Liu}}, \bibinfo {author} {\bibfnamefont {S.}~\bibnamefont {Jordan}}, \bibinfo {author} {\bibfnamefont {A.~V.}\ \bibnamefont {Gorshkov}},\ and\ \bibinfo {author} {\bibfnamefont
  {C.}~\bibnamefont {Monroe}},\ }\bibfield  {title} {\bibinfo {title} {Quantum approximate optimization of the long-range {I}sing model with a trapped-ion quantum simulator},\ }\href {https://www.pnas.org/content/117/41/25396} {\bibfield  {journal} {\bibinfo  {journal} {Proc. Natl. Acad. Sci.}\ }\textbf {\bibinfo {volume} {117}},\ \bibinfo {pages} {25396} (\bibinfo {year} {2020})}\BibitemShut {NoStop}%
\bibitem [{\citenamefont {Harrigan}\ \emph {et~al.}(2021)\citenamefont {Harrigan}, \citenamefont {Sung}, \citenamefont {Neeley}, \citenamefont {Satzinger}, \citenamefont {Arute}, \citenamefont {Arya}, \citenamefont {Atalaya}, \citenamefont {Bardin}, \citenamefont {Barends}, \citenamefont {Boixo} \emph {et~al.}}]{harrigan2021quantum}%
  \BibitemOpen
  \bibfield  {author} {\bibinfo {author} {\bibfnamefont {M.~P.}\ \bibnamefont {Harrigan}}, \bibinfo {author} {\bibfnamefont {K.~J.}\ \bibnamefont {Sung}}, \bibinfo {author} {\bibfnamefont {M.}~\bibnamefont {Neeley}}, \bibinfo {author} {\bibfnamefont {K.~J.}\ \bibnamefont {Satzinger}}, \bibinfo {author} {\bibfnamefont {F.}~\bibnamefont {Arute}}, \bibinfo {author} {\bibfnamefont {K.}~\bibnamefont {Arya}}, \bibinfo {author} {\bibfnamefont {J.}~\bibnamefont {Atalaya}}, \bibinfo {author} {\bibfnamefont {J.~C.}\ \bibnamefont {Bardin}}, \bibinfo {author} {\bibfnamefont {R.}~\bibnamefont {Barends}}, \bibinfo {author} {\bibfnamefont {S.}~\bibnamefont {Boixo}}, \emph {et~al.},\ }\bibfield  {title} {\bibinfo {title} {Quantum approximate optimization of non-planar graph problems on a planar superconducting processor},\ }\href@noop {} {\bibfield  {journal} {\bibinfo  {journal} {Nat. Phys.}\ }\textbf {\bibinfo {volume} {17}},\ \bibinfo {pages} {332} (\bibinfo {year} {2021})}\BibitemShut {NoStop}%
\bibitem [{\citenamefont {Zhu}\ \emph {et~al.}(2022)\citenamefont {Zhu}, \citenamefont {Zhang}, \citenamefont {Sundar}, \citenamefont {Green}, \citenamefont {Alderete}, \citenamefont {Nguyen}, \citenamefont {Hazzard},\ and\ \citenamefont {Linke}}]{Zhu2022Multi-roundComputer}%
  \BibitemOpen
  \bibfield  {author} {\bibinfo {author} {\bibfnamefont {Y.}~\bibnamefont {Zhu}}, \bibinfo {author} {\bibfnamefont {Z.}~\bibnamefont {Zhang}}, \bibinfo {author} {\bibfnamefont {B.}~\bibnamefont {Sundar}}, \bibinfo {author} {\bibfnamefont {A.~M.}\ \bibnamefont {Green}}, \bibinfo {author} {\bibfnamefont {C.~H.}\ \bibnamefont {Alderete}}, \bibinfo {author} {\bibfnamefont {N.~H.}\ \bibnamefont {Nguyen}}, \bibinfo {author} {\bibfnamefont {K.~R.}\ \bibnamefont {Hazzard}},\ and\ \bibinfo {author} {\bibfnamefont {N.~M.}\ \bibnamefont {Linke}},\ }\bibfield  {title} {\bibinfo {title} {Multi-round {QAOA} and advanced mixers on a trapped-ion quantum computer},\ }\href@noop {} {\bibfield  {journal} {\bibinfo  {journal} {Quantum Sci. Technol.}\ }\textbf {\bibinfo {volume} {8}},\ \bibinfo {pages} {015007} (\bibinfo {year} {2022})}\BibitemShut {NoStop}%
\bibitem [{\citenamefont {Moses}\ \emph {et~al.}(2023{\natexlab{a}})\citenamefont {Moses}, \citenamefont {Baldwin}, \citenamefont {Allman}, \citenamefont {Ancona}, \citenamefont {Ascarrunz}, \citenamefont {Barnes}, \citenamefont {Bartolotta}, \citenamefont {Bjork}, \citenamefont {Blanchard}, \citenamefont {Bohn} \emph {et~al.}}]{moses2023race}%
  \BibitemOpen
  \bibfield  {author} {\bibinfo {author} {\bibfnamefont {S.}~\bibnamefont {Moses}}, \bibinfo {author} {\bibfnamefont {C.}~\bibnamefont {Baldwin}}, \bibinfo {author} {\bibfnamefont {M.}~\bibnamefont {Allman}}, \bibinfo {author} {\bibfnamefont {R.}~\bibnamefont {Ancona}}, \bibinfo {author} {\bibfnamefont {L.}~\bibnamefont {Ascarrunz}}, \bibinfo {author} {\bibfnamefont {C.}~\bibnamefont {Barnes}}, \bibinfo {author} {\bibfnamefont {J.}~\bibnamefont {Bartolotta}}, \bibinfo {author} {\bibfnamefont {B.}~\bibnamefont {Bjork}}, \bibinfo {author} {\bibfnamefont {P.}~\bibnamefont {Blanchard}}, \bibinfo {author} {\bibfnamefont {M.}~\bibnamefont {Bohn}}, \emph {et~al.},\ }\bibfield  {title} {\bibinfo {title} {A race track trapped-ion quantum processor},\ }\href@noop {} {\bibfield  {journal} {\bibinfo  {journal} {arXiv preprint arXiv:2305.03828}\ } (\bibinfo {year} {2023}{\natexlab{a}})}\BibitemShut {NoStop}%
\bibitem [{\citenamefont {Shaydulin}\ and\ \citenamefont {Pistoia}(2023)}]{shaydulin2023qaoa}%
  \BibitemOpen
  \bibfield  {author} {\bibinfo {author} {\bibfnamefont {R.}~\bibnamefont {Shaydulin}}\ and\ \bibinfo {author} {\bibfnamefont {M.}~\bibnamefont {Pistoia}},\ }\bibfield  {title} {\bibinfo {title} {{QAOA} with {$N\cdot p\geq 200$}},\ }\href@noop {} {\bibfield  {journal} {\bibinfo  {journal} {arXiv preprint arXiv:2303.02064}\ } (\bibinfo {year} {2023})}\BibitemShut {NoStop}%
\bibitem [{\citenamefont {Shaydulin}\ \emph {et~al.}(2023{\natexlab{a}})\citenamefont {Shaydulin}, \citenamefont {Li}, \citenamefont {Chakrabarti}, \citenamefont {DeCross}, \citenamefont {Herman}, \citenamefont {Kumar}, \citenamefont {Larson}, \citenamefont {Lykov}, \citenamefont {Minssen}, \citenamefont {Sun} \emph {et~al.}}]{shaydulin2023evidence}%
  \BibitemOpen
  \bibfield  {author} {\bibinfo {author} {\bibfnamefont {R.}~\bibnamefont {Shaydulin}}, \bibinfo {author} {\bibfnamefont {C.}~\bibnamefont {Li}}, \bibinfo {author} {\bibfnamefont {S.}~\bibnamefont {Chakrabarti}}, \bibinfo {author} {\bibfnamefont {M.}~\bibnamefont {DeCross}}, \bibinfo {author} {\bibfnamefont {D.}~\bibnamefont {Herman}}, \bibinfo {author} {\bibfnamefont {N.}~\bibnamefont {Kumar}}, \bibinfo {author} {\bibfnamefont {J.}~\bibnamefont {Larson}}, \bibinfo {author} {\bibfnamefont {D.}~\bibnamefont {Lykov}}, \bibinfo {author} {\bibfnamefont {P.}~\bibnamefont {Minssen}}, \bibinfo {author} {\bibfnamefont {Y.}~\bibnamefont {Sun}}, \emph {et~al.},\ }\bibfield  {title} {\bibinfo {title} {Evidence of scaling advantage for the quantum approximate optimization algorithm on a classically intractable problem},\ }\href@noop {} {\bibfield  {journal} {\bibinfo  {journal} {arXiv preprint arXiv:2308.02342}\ } (\bibinfo {year} {2023}{\natexlab{a}})}\BibitemShut {NoStop}%
\bibitem [{\citenamefont {Lubinski}\ \emph {et~al.}(2023)\citenamefont {Lubinski}, \citenamefont {Coffrin}, \citenamefont {McGeoch}, \citenamefont {Sathe}, \citenamefont {Apanavicius},\ and\ \citenamefont {Neira}}]{lubinski2023optimization}%
  \BibitemOpen
  \bibfield  {author} {\bibinfo {author} {\bibfnamefont {T.}~\bibnamefont {Lubinski}}, \bibinfo {author} {\bibfnamefont {C.}~\bibnamefont {Coffrin}}, \bibinfo {author} {\bibfnamefont {C.}~\bibnamefont {McGeoch}}, \bibinfo {author} {\bibfnamefont {P.}~\bibnamefont {Sathe}}, \bibinfo {author} {\bibfnamefont {J.}~\bibnamefont {Apanavicius}},\ and\ \bibinfo {author} {\bibfnamefont {D.~E.~B.}\ \bibnamefont {Neira}},\ }\bibfield  {title} {\bibinfo {title} {Optimization applications as quantum performance benchmarks},\ }\href@noop {} {\bibfield  {journal} {\bibinfo  {journal} {arXiv preprint arXiv:2302.02278}\ } (\bibinfo {year} {2023})}\BibitemShut {NoStop}%
\bibitem [{\citenamefont {Dupont}\ \emph {et~al.}(2023)\citenamefont {Dupont}, \citenamefont {Evert}, \citenamefont {Hodson}, \citenamefont {Sundar}, \citenamefont {Jeffrey}, \citenamefont {Yamaguchi}, \citenamefont {Feng}, \citenamefont {Maciejewski}, \citenamefont {Hadfield}, \citenamefont {Alam} \emph {et~al.}}]{dupont2023quantum}%
  \BibitemOpen
  \bibfield  {author} {\bibinfo {author} {\bibfnamefont {M.}~\bibnamefont {Dupont}}, \bibinfo {author} {\bibfnamefont {B.}~\bibnamefont {Evert}}, \bibinfo {author} {\bibfnamefont {M.~J.}\ \bibnamefont {Hodson}}, \bibinfo {author} {\bibfnamefont {B.}~\bibnamefont {Sundar}}, \bibinfo {author} {\bibfnamefont {S.}~\bibnamefont {Jeffrey}}, \bibinfo {author} {\bibfnamefont {Y.}~\bibnamefont {Yamaguchi}}, \bibinfo {author} {\bibfnamefont {D.}~\bibnamefont {Feng}}, \bibinfo {author} {\bibfnamefont {F.~B.}\ \bibnamefont {Maciejewski}}, \bibinfo {author} {\bibfnamefont {S.}~\bibnamefont {Hadfield}}, \bibinfo {author} {\bibfnamefont {M.~S.}\ \bibnamefont {Alam}}, \emph {et~al.},\ }\bibfield  {title} {\bibinfo {title} {Quantum enhanced greedy solver for optimization problems},\ }\href@noop {} {\bibfield  {journal} {\bibinfo  {journal} {arXiv preprint arXiv:2303.05509}\ } (\bibinfo {year} {2023})}\BibitemShut {NoStop}%
\bibitem [{\citenamefont {Maciejewski}\ \emph {et~al.}(2023)\citenamefont {Maciejewski}, \citenamefont {Hadfield}, \citenamefont {Hall}, \citenamefont {Hodson}, \citenamefont {Dupont}, \citenamefont {Evert}, \citenamefont {Sud}, \citenamefont {Alam}, \citenamefont {Wang}, \citenamefont {Jeffrey} \emph {et~al.}}]{maciejewski2023design}%
  \BibitemOpen
  \bibfield  {author} {\bibinfo {author} {\bibfnamefont {F.~B.}\ \bibnamefont {Maciejewski}}, \bibinfo {author} {\bibfnamefont {S.}~\bibnamefont {Hadfield}}, \bibinfo {author} {\bibfnamefont {B.}~\bibnamefont {Hall}}, \bibinfo {author} {\bibfnamefont {M.}~\bibnamefont {Hodson}}, \bibinfo {author} {\bibfnamefont {M.}~\bibnamefont {Dupont}}, \bibinfo {author} {\bibfnamefont {B.}~\bibnamefont {Evert}}, \bibinfo {author} {\bibfnamefont {J.}~\bibnamefont {Sud}}, \bibinfo {author} {\bibfnamefont {M.~S.}\ \bibnamefont {Alam}}, \bibinfo {author} {\bibfnamefont {Z.}~\bibnamefont {Wang}}, \bibinfo {author} {\bibfnamefont {S.}~\bibnamefont {Jeffrey}}, \emph {et~al.},\ }\bibfield  {title} {\bibinfo {title} {Design and execution of quantum circuits using tens of superconducting qubits and thousands of gates for dense {Ising} optimization problems},\ }\href@noop {} {\bibfield  {journal} {\bibinfo  {journal} {arXiv preprint arXiv:2308.12423}\ } (\bibinfo {year} {2023})}\BibitemShut {NoStop}%
\bibitem [{\citenamefont {Wurtz}\ and\ \citenamefont {Lykov}(2021)}]{wurtz2021fixed}%
  \BibitemOpen
  \bibfield  {author} {\bibinfo {author} {\bibfnamefont {J.}~\bibnamefont {Wurtz}}\ and\ \bibinfo {author} {\bibfnamefont {D.}~\bibnamefont {Lykov}},\ }\bibfield  {title} {\bibinfo {title} {Fixed-angle conjectures for the quantum approximate optimization algorithm on regular {M}ax{C}ut graphs},\ }\href@noop {} {\bibfield  {journal} {\bibinfo  {journal} {Phys. Rev. A}\ }\textbf {\bibinfo {volume} {104}},\ \bibinfo {pages} {052419} (\bibinfo {year} {2021})}\BibitemShut {NoStop}%
\bibitem [{\citenamefont {Kremenetski}\ \emph {et~al.}(2023)\citenamefont {Kremenetski}, \citenamefont {Apte}, \citenamefont {Hogg}, \citenamefont {Hadfield},\ and\ \citenamefont {Tubman}}]{kremenetski2023quantum}%
  \BibitemOpen
  \bibfield  {author} {\bibinfo {author} {\bibfnamefont {V.}~\bibnamefont {Kremenetski}}, \bibinfo {author} {\bibfnamefont {A.}~\bibnamefont {Apte}}, \bibinfo {author} {\bibfnamefont {T.}~\bibnamefont {Hogg}}, \bibinfo {author} {\bibfnamefont {S.}~\bibnamefont {Hadfield}},\ and\ \bibinfo {author} {\bibfnamefont {N.~M.}\ \bibnamefont {Tubman}},\ }\bibfield  {title} {\bibinfo {title} {Quantum alternating operator ansatz ({QAOA}) beyond low depth with gradually changing unitaries},\ }\href@noop {} {\bibfield  {journal} {\bibinfo  {journal} {arXiv preprint arXiv:2305.04455}\ } (\bibinfo {year} {2023})}\BibitemShut {NoStop}%
\bibitem [{\citenamefont {Yang}\ \emph {et~al.}(2017)\citenamefont {Yang}, \citenamefont {Rahmani}, \citenamefont {Shabani}, \citenamefont {Neven},\ and\ \citenamefont {Chamon}}]{yang2017optimizing}%
  \BibitemOpen
  \bibfield  {author} {\bibinfo {author} {\bibfnamefont {Z.-C.}\ \bibnamefont {Yang}}, \bibinfo {author} {\bibfnamefont {A.}~\bibnamefont {Rahmani}}, \bibinfo {author} {\bibfnamefont {A.}~\bibnamefont {Shabani}}, \bibinfo {author} {\bibfnamefont {H.}~\bibnamefont {Neven}},\ and\ \bibinfo {author} {\bibfnamefont {C.}~\bibnamefont {Chamon}},\ }\bibfield  {title} {\bibinfo {title} {Optimizing variational quantum algorithms using {P}ontryagin’s minimum principle},\ }\href@noop {} {\bibfield  {journal} {\bibinfo  {journal} {Phys. Rev. X}\ }\textbf {\bibinfo {volume} {7}},\ \bibinfo {pages} {021027} (\bibinfo {year} {2017})}\BibitemShut {NoStop}%
\bibitem [{\citenamefont {Sack}\ and\ \citenamefont {Serbyn}(2021)}]{sack2021quantum}%
  \BibitemOpen
  \bibfield  {author} {\bibinfo {author} {\bibfnamefont {S.~H.}\ \bibnamefont {Sack}}\ and\ \bibinfo {author} {\bibfnamefont {M.}~\bibnamefont {Serbyn}},\ }\bibfield  {title} {\bibinfo {title} {Quantum annealing initialization of the quantum approximate optimization algorithm},\ }\href@noop {} {\bibfield  {journal} {\bibinfo  {journal} {Quantum}\ }\textbf {\bibinfo {volume} {5}},\ \bibinfo {pages} {491} (\bibinfo {year} {2021})}\BibitemShut {NoStop}%
\bibitem [{\citenamefont {Brady}\ \emph {et~al.}(2021)\citenamefont {Brady}, \citenamefont {Baldwin}, \citenamefont {Bapat}, \citenamefont {Kharkov},\ and\ \citenamefont {Gorshkov}}]{brady2021optimal}%
  \BibitemOpen
  \bibfield  {author} {\bibinfo {author} {\bibfnamefont {L.~T.}\ \bibnamefont {Brady}}, \bibinfo {author} {\bibfnamefont {C.~L.}\ \bibnamefont {Baldwin}}, \bibinfo {author} {\bibfnamefont {A.}~\bibnamefont {Bapat}}, \bibinfo {author} {\bibfnamefont {Y.}~\bibnamefont {Kharkov}},\ and\ \bibinfo {author} {\bibfnamefont {A.~V.}\ \bibnamefont {Gorshkov}},\ }\bibfield  {title} {\bibinfo {title} {Optimal protocols in quantum annealing and quantum approximate optimization algorithm problems},\ }\href@noop {} {\bibfield  {journal} {\bibinfo  {journal} {Phys. Rev. Lett.}\ }\textbf {\bibinfo {volume} {126}},\ \bibinfo {pages} {070505} (\bibinfo {year} {2021})}\BibitemShut {NoStop}%
\bibitem [{\citenamefont {Wurtz}\ and\ \citenamefont {Love}(2022)}]{wurtz2022counterdiabaticity}%
  \BibitemOpen
  \bibfield  {author} {\bibinfo {author} {\bibfnamefont {J.}~\bibnamefont {Wurtz}}\ and\ \bibinfo {author} {\bibfnamefont {P.~J.}\ \bibnamefont {Love}},\ }\bibfield  {title} {\bibinfo {title} {Counterdiabaticity and the quantum approximate optimization algorithm},\ }\href@noop {} {\bibfield  {journal} {\bibinfo  {journal} {Quantum}\ }\textbf {\bibinfo {volume} {6}},\ \bibinfo {pages} {635} (\bibinfo {year} {2022})}\BibitemShut {NoStop}%
\bibitem [{\citenamefont {Wu}\ \emph {et~al.}(2023)\citenamefont {Wu}, \citenamefont {Liu},\ and\ \citenamefont {Chen}}]{wu2023adiabaticpassage}%
  \BibitemOpen
  \bibfield  {author} {\bibinfo {author} {\bibfnamefont {M.}~\bibnamefont {Wu}}, \bibinfo {author} {\bibfnamefont {Z.}~\bibnamefont {Liu}},\ and\ \bibinfo {author} {\bibfnamefont {H.}~\bibnamefont {Chen}},\ }\bibfield  {title} {\bibinfo {title} {Adiabatic-passage based parameter setting method for quantum approximate optimization algorithm on 3-{SAT} problem},\ }\href@noop {} {\bibfield  {journal} {\bibinfo  {journal} {arXiv preprint arXiv:2312.00077}\ } (\bibinfo {year} {2023})}\BibitemShut {NoStop}%
\bibitem [{\citenamefont {Streif}\ and\ \citenamefont {Leib}(2020)}]{streif2020training}%
  \BibitemOpen
  \bibfield  {author} {\bibinfo {author} {\bibfnamefont {M.}~\bibnamefont {Streif}}\ and\ \bibinfo {author} {\bibfnamefont {M.}~\bibnamefont {Leib}},\ }\bibfield  {title} {\bibinfo {title} {Training the quantum approximate optimization algorithm without access to a quantum processing unit},\ }\href@noop {} {\bibfield  {journal} {\bibinfo  {journal} {Quantum Sci. Technol.}\ }\textbf {\bibinfo {volume} {5}},\ \bibinfo {pages} {034008} (\bibinfo {year} {2020})}\BibitemShut {NoStop}%
\bibitem [{\citenamefont {Moussa}\ \emph {et~al.}(2022)\citenamefont {Moussa}, \citenamefont {Wang}, \citenamefont {B{\"a}ck},\ and\ \citenamefont {Dunjko}}]{moussa2022unsupervised}%
  \BibitemOpen
  \bibfield  {author} {\bibinfo {author} {\bibfnamefont {C.}~\bibnamefont {Moussa}}, \bibinfo {author} {\bibfnamefont {H.}~\bibnamefont {Wang}}, \bibinfo {author} {\bibfnamefont {T.}~\bibnamefont {B{\"a}ck}},\ and\ \bibinfo {author} {\bibfnamefont {V.}~\bibnamefont {Dunjko}},\ }\bibfield  {title} {\bibinfo {title} {Unsupervised strategies for identifying optimal parameters in quantum approximate optimization algorithm},\ }\href@noop {} {\bibfield  {journal} {\bibinfo  {journal} {EPJ Quantum Technol.}\ }\textbf {\bibinfo {volume} {9}},\ \bibinfo {pages} {11} (\bibinfo {year} {2022})}\BibitemShut {NoStop}%
\bibitem [{\citenamefont {Farhi}\ \emph {et~al.}(2022)\citenamefont {Farhi}, \citenamefont {Goldstone}, \citenamefont {Gutmann},\ and\ \citenamefont {Zhou}}]{farhi2022quantum}%
  \BibitemOpen
  \bibfield  {author} {\bibinfo {author} {\bibfnamefont {E.}~\bibnamefont {Farhi}}, \bibinfo {author} {\bibfnamefont {J.}~\bibnamefont {Goldstone}}, \bibinfo {author} {\bibfnamefont {S.}~\bibnamefont {Gutmann}},\ and\ \bibinfo {author} {\bibfnamefont {L.}~\bibnamefont {Zhou}},\ }\bibfield  {title} {\bibinfo {title} {The quantum approximate optimization algorithm and the {S}herrington-{K}irkpatrick model at infinite size},\ }\href@noop {} {\bibfield  {journal} {\bibinfo  {journal} {Quantum}\ }\textbf {\bibinfo {volume} {6}},\ \bibinfo {pages} {759} (\bibinfo {year} {2022})}\BibitemShut {NoStop}%
\bibitem [{\citenamefont {Galda}\ \emph {et~al.}(2023)\citenamefont {Galda}, \citenamefont {Gupta}, \citenamefont {Falla}, \citenamefont {Liu}, \citenamefont {Lykov}, \citenamefont {Alexeev},\ and\ \citenamefont {Safro}}]{galda2023similarity}%
  \BibitemOpen
  \bibfield  {author} {\bibinfo {author} {\bibfnamefont {A.}~\bibnamefont {Galda}}, \bibinfo {author} {\bibfnamefont {E.}~\bibnamefont {Gupta}}, \bibinfo {author} {\bibfnamefont {J.}~\bibnamefont {Falla}}, \bibinfo {author} {\bibfnamefont {X.}~\bibnamefont {Liu}}, \bibinfo {author} {\bibfnamefont {D.}~\bibnamefont {Lykov}}, \bibinfo {author} {\bibfnamefont {Y.}~\bibnamefont {Alexeev}},\ and\ \bibinfo {author} {\bibfnamefont {I.}~\bibnamefont {Safro}},\ }\bibfield  {title} {\bibinfo {title} {Similarity-based parameter transferability in the quantum approximate optimization algorithm},\ }\href@noop {} {\bibfield  {journal} {\bibinfo  {journal} {arXiv preprint arXiv:2307.05420}\ } (\bibinfo {year} {2023})}\BibitemShut {NoStop}%
\bibitem [{\citenamefont {Shaydulin}\ \emph {et~al.}(2023{\natexlab{b}})\citenamefont {Shaydulin}, \citenamefont {Lotshaw}, \citenamefont {Larson}, \citenamefont {Ostrowski},\ and\ \citenamefont {Humble}}]{shaydulin2023parameter}%
  \BibitemOpen
  \bibfield  {author} {\bibinfo {author} {\bibfnamefont {R.}~\bibnamefont {Shaydulin}}, \bibinfo {author} {\bibfnamefont {P.~C.}\ \bibnamefont {Lotshaw}}, \bibinfo {author} {\bibfnamefont {J.}~\bibnamefont {Larson}}, \bibinfo {author} {\bibfnamefont {J.}~\bibnamefont {Ostrowski}},\ and\ \bibinfo {author} {\bibfnamefont {T.~S.}\ \bibnamefont {Humble}},\ }\bibfield  {title} {\bibinfo {title} {Parameter transfer for quantum approximate optimization of weighted {M}ax{C}ut},\ }\href@noop {} {\bibfield  {journal} {\bibinfo  {journal} {ACM Transactions on Quantum Computing}\ }\textbf {\bibinfo {volume} {4}},\ \bibinfo {pages} {1} (\bibinfo {year} {2023}{\natexlab{b}})}\BibitemShut {NoStop}%
\bibitem [{\citenamefont {Kapit}\ \emph {et~al.}(2023)\citenamefont {Kapit}, \citenamefont {Barton}, \citenamefont {Feeney}, \citenamefont {Grattan}, \citenamefont {Patnaik}, \citenamefont {Sagal}, \citenamefont {Carr},\ and\ \citenamefont {Oganesyan}}]{kapit2023approximability}%
  \BibitemOpen
  \bibfield  {author} {\bibinfo {author} {\bibfnamefont {E.}~\bibnamefont {Kapit}}, \bibinfo {author} {\bibfnamefont {B.~A.}\ \bibnamefont {Barton}}, \bibinfo {author} {\bibfnamefont {S.}~\bibnamefont {Feeney}}, \bibinfo {author} {\bibfnamefont {G.}~\bibnamefont {Grattan}}, \bibinfo {author} {\bibfnamefont {P.}~\bibnamefont {Patnaik}}, \bibinfo {author} {\bibfnamefont {J.}~\bibnamefont {Sagal}}, \bibinfo {author} {\bibfnamefont {L.~D.}\ \bibnamefont {Carr}},\ and\ \bibinfo {author} {\bibfnamefont {V.}~\bibnamefont {Oganesyan}},\ }\bibfield  {title} {\bibinfo {title} {On the approximability of random-hypergraph {MAX-3-XORSAT} problems with quantum algorithms},\ }\href@noop {} {\bibfield  {journal} {\bibinfo  {journal} {arXiv preprint arXiv:2312.06104}\ } (\bibinfo {year} {2023})}\BibitemShut {NoStop}%
\bibitem [{\citenamefont {Yu}\ \emph {et~al.}(2023)\citenamefont {Yu}, \citenamefont {Cao}, \citenamefont {Wang}, \citenamefont {Shannon},\ and\ \citenamefont {Joynt}}]{yu2023solution}%
  \BibitemOpen
  \bibfield  {author} {\bibinfo {author} {\bibfnamefont {Y.}~\bibnamefont {Yu}}, \bibinfo {author} {\bibfnamefont {C.}~\bibnamefont {Cao}}, \bibinfo {author} {\bibfnamefont {X.-B.}\ \bibnamefont {Wang}}, \bibinfo {author} {\bibfnamefont {N.}~\bibnamefont {Shannon}},\ and\ \bibinfo {author} {\bibfnamefont {R.}~\bibnamefont {Joynt}},\ }\bibfield  {title} {\bibinfo {title} {Solution of {SAT} problems with the adaptive-bias quantum approximate optimization algorithm},\ }\href@noop {} {\bibfield  {journal} {\bibinfo  {journal} {Phys. Rev. Res.}\ }\textbf {\bibinfo {volume} {5}},\ \bibinfo {pages} {023147} (\bibinfo {year} {2023})}\BibitemShut {NoStop}%
\bibitem [{\citenamefont {Cook}(2023)}]{cook2023complexity}%
  \BibitemOpen
  \bibfield  {author} {\bibinfo {author} {\bibfnamefont {S.~A.}\ \bibnamefont {Cook}},\ }\bibfield  {title} {\bibinfo {title} {The complexity of theorem-proving procedures},\ }in\ \href@noop {} {\emph {\bibinfo {booktitle} {Logic, Automata, and Computational Complexity: The Works of Stephen A. Cook}}}\ (\bibinfo {year} {2023})\ pp.\ \bibinfo {pages} {143--152}\BibitemShut {NoStop}%
\bibitem [{\citenamefont {Karp}(2010)}]{karp2010reducibility}%
  \BibitemOpen
  \bibfield  {author} {\bibinfo {author} {\bibfnamefont {R.~M.}\ \bibnamefont {Karp}},\ }\bibinfo {title} {Reducibility among combinatorial problems},\ in\ \href@noop {} {\emph {\bibinfo {booktitle} {50 Years of Integer Programming 1958-2008: From the Early Years to the State-of-the-Art}}},\ \bibinfo {editor} {edited by\ \bibinfo {editor} {\bibfnamefont {M.}~\bibnamefont {J{\"u}nger}}, \bibinfo {editor} {\bibfnamefont {T.~M.}\ \bibnamefont {Liebling}}, \bibinfo {editor} {\bibfnamefont {D.}~\bibnamefont {Naddef}}, \bibinfo {editor} {\bibfnamefont {G.~L.}\ \bibnamefont {Nemhauser}}, \bibinfo {editor} {\bibfnamefont {W.~R.}\ \bibnamefont {Pulleyblank}}, \bibinfo {editor} {\bibfnamefont {G.}~\bibnamefont {Reinelt}}, \bibinfo {editor} {\bibfnamefont {G.}~\bibnamefont {Rinaldi}},\ and\ \bibinfo {editor} {\bibfnamefont {L.~A.}\ \bibnamefont {Wolsey}}}\ (\bibinfo  {publisher} {Springer Berlin Heidelberg},\ \bibinfo {address} {Berlin, Heidelberg},\ \bibinfo {year} {2010})\ pp.\ \bibinfo {pages} {219--241}\BibitemShut
  {NoStop}%
\bibitem [{\citenamefont {Lucas}(2014)}]{lucas2014ising}%
  \BibitemOpen
  \bibfield  {author} {\bibinfo {author} {\bibfnamefont {A.}~\bibnamefont {Lucas}},\ }\bibfield  {title} {\bibinfo {title} {Ising formulations of many {NP} problems},\ }\href@noop {} {\bibfield  {journal} {\bibinfo  {journal} {Front. Phys.}\ }\textbf {\bibinfo {volume} {2}},\ \bibinfo {pages} {5} (\bibinfo {year} {2014})}\BibitemShut {NoStop}%
\bibitem [{\citenamefont {Khurshid}\ and\ \citenamefont {Marinov}(2004)}]{khurshid2004testera}%
  \BibitemOpen
  \bibfield  {author} {\bibinfo {author} {\bibfnamefont {S.}~\bibnamefont {Khurshid}}\ and\ \bibinfo {author} {\bibfnamefont {D.}~\bibnamefont {Marinov}},\ }\bibfield  {title} {\bibinfo {title} {Test{E}ra: {S}pecification-based testing of {J}ava programs using {SAT}},\ }\href@noop {} {\bibfield  {journal} {\bibinfo  {journal} {Autom. Softw. Eng.}\ }\textbf {\bibinfo {volume} {11}},\ \bibinfo {pages} {403} (\bibinfo {year} {2004})}\BibitemShut {NoStop}%
\bibitem [{\citenamefont {Vizel}\ \emph {et~al.}(2015)\citenamefont {Vizel}, \citenamefont {Weissenbacher},\ and\ \citenamefont {Malik}}]{Vizel2015BooleanChecking}%
  \BibitemOpen
  \bibfield  {author} {\bibinfo {author} {\bibfnamefont {Y.}~\bibnamefont {Vizel}}, \bibinfo {author} {\bibfnamefont {G.}~\bibnamefont {Weissenbacher}},\ and\ \bibinfo {author} {\bibfnamefont {S.}~\bibnamefont {Malik}},\ }\bibfield  {title} {\bibinfo {title} {{B}oolean satisfiability solvers and their applications in model checking},\ }\href@noop {} {\bibfield  {journal} {\bibinfo  {journal} {Proc. IEEE}\ }\textbf {\bibinfo {volume} {103}},\ \bibinfo {pages} {2021} (\bibinfo {year} {2015})}\BibitemShut {NoStop}%
\bibitem [{\citenamefont {Dutra}\ \emph {et~al.}(2018)\citenamefont {Dutra}, \citenamefont {Laeufer}, \citenamefont {Bachrach},\ and\ \citenamefont {Sen}}]{Dutra2018EfficientTesting}%
  \BibitemOpen
  \bibfield  {author} {\bibinfo {author} {\bibfnamefont {R.}~\bibnamefont {Dutra}}, \bibinfo {author} {\bibfnamefont {K.}~\bibnamefont {Laeufer}}, \bibinfo {author} {\bibfnamefont {J.}~\bibnamefont {Bachrach}},\ and\ \bibinfo {author} {\bibfnamefont {K.}~\bibnamefont {Sen}},\ }\bibfield  {title} {\bibinfo {title} {{Efficient sampling of SAT solutions for testing}},\ }in\ \href@noop {} {\emph {\bibinfo {booktitle} {Proceedings of the 40th International Conference on Software Engineering}}}\ (\bibinfo  {publisher} {ACM},\ \bibinfo {address} {New York, NY, USA},\ \bibinfo {year} {2018})\ pp.\ \bibinfo {pages} {549--559}\BibitemShut {NoStop}%
\bibitem [{\citenamefont {Gaber}\ \emph {et~al.}(2020)\citenamefont {Gaber}, \citenamefont {Hussein}, \citenamefont {Mahmoud}, \citenamefont {Mabrook},\ and\ \citenamefont {Moness}}]{gaber2020computation}%
  \BibitemOpen
  \bibfield  {author} {\bibinfo {author} {\bibfnamefont {L.}~\bibnamefont {Gaber}}, \bibinfo {author} {\bibfnamefont {A.~I.}\ \bibnamefont {Hussein}}, \bibinfo {author} {\bibfnamefont {H.}~\bibnamefont {Mahmoud}}, \bibinfo {author} {\bibfnamefont {M.~M.}\ \bibnamefont {Mabrook}},\ and\ \bibinfo {author} {\bibfnamefont {M.}~\bibnamefont {Moness}},\ }\bibfield  {title} {\bibinfo {title} {Computation of minimal unsatisfiable subformulas for {SAT}-based digital circuit error diagnosis},\ }\href@noop {} {\bibfield  {journal} {\bibinfo  {journal} {J. Ambient. Intell. Humaniz. Comput.}\ ,\ \bibinfo {pages} {1}} (\bibinfo {year} {2020})}\BibitemShut {NoStop}%
\bibitem [{\citenamefont {Lafitte}\ \emph {et~al.}(2014)\citenamefont {Lafitte}, \citenamefont {Nakahara~Jr},\ and\ \citenamefont {Van~Heule}}]{lafitte2014applications}%
  \BibitemOpen
  \bibfield  {author} {\bibinfo {author} {\bibfnamefont {F.}~\bibnamefont {Lafitte}}, \bibinfo {author} {\bibfnamefont {J.}~\bibnamefont {Nakahara~Jr}},\ and\ \bibinfo {author} {\bibfnamefont {D.}~\bibnamefont {Van~Heule}},\ }\bibfield  {title} {\bibinfo {title} {Applications of {SAT} solvers in cryptanalysis: finding weak keys and preimages},\ }\href@noop {} {\bibfield  {journal} {\bibinfo  {journal} {J. Satisf. Boolean Model. Comput.}\ }\textbf {\bibinfo {volume} {9}},\ \bibinfo {pages} {1} (\bibinfo {year} {2014})}\BibitemShut {NoStop}%
\bibitem [{\citenamefont {Lynce}\ and\ \citenamefont {Marques-Silva}(2006)}]{lynce2006efficient}%
  \BibitemOpen
  \bibfield  {author} {\bibinfo {author} {\bibfnamefont {I.}~\bibnamefont {Lynce}}\ and\ \bibinfo {author} {\bibfnamefont {J.}~\bibnamefont {Marques-Silva}},\ }\bibfield  {title} {\bibinfo {title} {Efficient haplotype inference with {B}oolean satisfiability},\ }in\ \href@noop {} {\emph {\bibinfo {booktitle} {National Conference on Artificial Intelligence (AAAI)}}}\ (\bibinfo {organization} {AAAI Press},\ \bibinfo {year} {2006})\BibitemShut {NoStop}%
\bibitem [{\citenamefont {Paredes}\ \emph {et~al.}(2019)\citenamefont {Paredes}, \citenamefont {Due{\~n}as-Osorio}, \citenamefont {Meel},\ and\ \citenamefont {Vardi}}]{paredes2019principled}%
  \BibitemOpen
  \bibfield  {author} {\bibinfo {author} {\bibfnamefont {R.}~\bibnamefont {Paredes}}, \bibinfo {author} {\bibfnamefont {L.}~\bibnamefont {Due{\~n}as-Osorio}}, \bibinfo {author} {\bibfnamefont {K.~S.}\ \bibnamefont {Meel}},\ and\ \bibinfo {author} {\bibfnamefont {M.~Y.}\ \bibnamefont {Vardi}},\ }\bibfield  {title} {\bibinfo {title} {Principled network reliability approximation: A counting-based approach},\ }\href@noop {} {\bibfield  {journal} {\bibinfo  {journal} {Reliability Engineering \& System Safety}\ }\textbf {\bibinfo {volume} {191}},\ \bibinfo {pages} {106472} (\bibinfo {year} {2019})}\BibitemShut {NoStop}%
\bibitem [{\citenamefont {Marques-Silva}(2008)}]{marques2008practical}%
  \BibitemOpen
  \bibfield  {author} {\bibinfo {author} {\bibfnamefont {J.}~\bibnamefont {Marques-Silva}},\ }\bibfield  {title} {\bibinfo {title} {Practical applications of {B}oolean satisfiability},\ }in\ \href@noop {} {\emph {\bibinfo {booktitle} {2008 9th International Workshop on Discrete Event Systems}}}\ (\bibinfo {organization} {IEEE},\ \bibinfo {year} {2008})\ pp.\ \bibinfo {pages} {74--80}\BibitemShut {NoStop}%
\bibitem [{\citenamefont {Yu}\ \emph {et~al.}(2014)\citenamefont {Yu}, \citenamefont {Subramanyan}, \citenamefont {Tsiskaridze},\ and\ \citenamefont {Malik}}]{yu2014all}%
  \BibitemOpen
  \bibfield  {author} {\bibinfo {author} {\bibfnamefont {Y.}~\bibnamefont {Yu}}, \bibinfo {author} {\bibfnamefont {P.}~\bibnamefont {Subramanyan}}, \bibinfo {author} {\bibfnamefont {N.}~\bibnamefont {Tsiskaridze}},\ and\ \bibinfo {author} {\bibfnamefont {S.}~\bibnamefont {Malik}},\ }\bibfield  {title} {\bibinfo {title} {All-{SAT} using minimal blocking clauses},\ }in\ \href@noop {} {\emph {\bibinfo {booktitle} {2014 27th International Conference on VLSI Design and 2014 13th International Conference on Embedded Systems}}}\ (\bibinfo {organization} {IEEE},\ \bibinfo {year} {2014})\ pp.\ \bibinfo {pages} {86--91}\BibitemShut {NoStop}%
\bibitem [{\citenamefont {Valiant}(1979)}]{valiant1979complexity}%
  \BibitemOpen
  \bibfield  {author} {\bibinfo {author} {\bibfnamefont {L.~G.}\ \bibnamefont {Valiant}},\ }\bibfield  {title} {\bibinfo {title} {The complexity of computing the permanent},\ }\href@noop {} {\bibfield  {journal} {\bibinfo  {journal} {Theor. Comput. Sci.}\ }\textbf {\bibinfo {volume} {8}},\ \bibinfo {pages} {189} (\bibinfo {year} {1979})}\BibitemShut {NoStop}%
\bibitem [{\citenamefont {Battaglia}\ \emph {et~al.}(2005)\citenamefont {Battaglia}, \citenamefont {Santoro},\ and\ \citenamefont {Tosatti}}]{battaglia2005optimization}%
  \BibitemOpen
  \bibfield  {author} {\bibinfo {author} {\bibfnamefont {D.~A.}\ \bibnamefont {Battaglia}}, \bibinfo {author} {\bibfnamefont {G.~E.}\ \bibnamefont {Santoro}},\ and\ \bibinfo {author} {\bibfnamefont {E.}~\bibnamefont {Tosatti}},\ }\bibfield  {title} {\bibinfo {title} {Optimization by quantum annealing: Lessons from hard satisfiability problems},\ }\href@noop {} {\bibfield  {journal} {\bibinfo  {journal} {Phys. Rev. E}\ }\textbf {\bibinfo {volume} {71}},\ \bibinfo {pages} {066707} (\bibinfo {year} {2005})}\BibitemShut {NoStop}%
\bibitem [{\citenamefont {Azinovi{\'c}}\ \emph {et~al.}(2017)\citenamefont {Azinovi{\'c}}, \citenamefont {Herr}, \citenamefont {Heim}, \citenamefont {Brown},\ and\ \citenamefont {Troyer}}]{azinovic2017assessment}%
  \BibitemOpen
  \bibfield  {author} {\bibinfo {author} {\bibfnamefont {M.}~\bibnamefont {Azinovi{\'c}}}, \bibinfo {author} {\bibfnamefont {D.}~\bibnamefont {Herr}}, \bibinfo {author} {\bibfnamefont {B.}~\bibnamefont {Heim}}, \bibinfo {author} {\bibfnamefont {E.}~\bibnamefont {Brown}},\ and\ \bibinfo {author} {\bibfnamefont {M.}~\bibnamefont {Troyer}},\ }\bibfield  {title} {\bibinfo {title} {Assessment of quantum annealing for the construction of satisfiability filters},\ }\href@noop {} {\bibfield  {journal} {\bibinfo  {journal} {SciPost Phys.}\ }\textbf {\bibinfo {volume} {2}},\ \bibinfo {pages} {013} (\bibinfo {year} {2017})}\BibitemShut {NoStop}%
\bibitem [{\citenamefont {Ayanzadeh}\ \emph {et~al.}(2020)\citenamefont {Ayanzadeh}, \citenamefont {Halem},\ and\ \citenamefont {Finin}}]{ayanzadeh2020reinforcement}%
  \BibitemOpen
  \bibfield  {author} {\bibinfo {author} {\bibfnamefont {R.}~\bibnamefont {Ayanzadeh}}, \bibinfo {author} {\bibfnamefont {M.}~\bibnamefont {Halem}},\ and\ \bibinfo {author} {\bibfnamefont {T.}~\bibnamefont {Finin}},\ }\bibfield  {title} {\bibinfo {title} {Reinforcement quantum annealing: A hybrid quantum learning automata},\ }\href@noop {} {\bibfield  {journal} {\bibinfo  {journal} {Sci. Rep.}\ }\textbf {\bibinfo {volume} {10}},\ \bibinfo {pages} {7952} (\bibinfo {year} {2020})}\BibitemShut {NoStop}%
\bibitem [{\citenamefont {Cheng}\ and\ \citenamefont {Tao}(2007)}]{cheng2007quantum}%
  \BibitemOpen
  \bibfield  {author} {\bibinfo {author} {\bibfnamefont {S.-T.}\ \bibnamefont {Cheng}}\ and\ \bibinfo {author} {\bibfnamefont {M.-H.}\ \bibnamefont {Tao}},\ }\bibfield  {title} {\bibinfo {title} {Quantum cooperative search algorithm for 3-{SAT}},\ }\href@noop {} {\bibfield  {journal} {\bibinfo  {journal} {J. Comput. Syst. Sci.}\ }\textbf {\bibinfo {volume} {73}},\ \bibinfo {pages} {123} (\bibinfo {year} {2007})}\BibitemShut {NoStop}%
\bibitem [{\citenamefont {Alasow}\ \emph {et~al.}(2022)\citenamefont {Alasow}, \citenamefont {Jin},\ and\ \citenamefont {Perkowski}}]{alasow2022quantum}%
  \BibitemOpen
  \bibfield  {author} {\bibinfo {author} {\bibfnamefont {A.}~\bibnamefont {Alasow}}, \bibinfo {author} {\bibfnamefont {P.}~\bibnamefont {Jin}},\ and\ \bibinfo {author} {\bibfnamefont {M.}~\bibnamefont {Perkowski}},\ }\bibfield  {title} {\bibinfo {title} {Quantum algorithm for variant maximum satisfiability},\ }\href@noop {} {\bibfield  {journal} {\bibinfo  {journal} {Entropy}\ }\textbf {\bibinfo {volume} {24}},\ \bibinfo {pages} {1615} (\bibinfo {year} {2022})}\BibitemShut {NoStop}%
\bibitem [{\citenamefont {Varmantchaonala}\ \emph {et~al.}(2023)\citenamefont {Varmantchaonala}, \citenamefont {Fendji}, \citenamefont {Njafa},\ and\ \citenamefont {Atemkeng}}]{varmantchaonala2023quantum}%
  \BibitemOpen
  \bibfield  {author} {\bibinfo {author} {\bibfnamefont {C.~M.}\ \bibnamefont {Varmantchaonala}}, \bibinfo {author} {\bibfnamefont {J.~L. K.~E.}\ \bibnamefont {Fendji}}, \bibinfo {author} {\bibfnamefont {J.~P.~T.}\ \bibnamefont {Njafa}},\ and\ \bibinfo {author} {\bibfnamefont {M.}~\bibnamefont {Atemkeng}},\ }\bibfield  {title} {\bibinfo {title} {Quantum hybrid algorithm for solving sat problem},\ }\href@noop {} {\bibfield  {journal} {\bibinfo  {journal} {Eng. Appl. Artif. Intell.}\ }\textbf {\bibinfo {volume} {121}},\ \bibinfo {pages} {106058} (\bibinfo {year} {2023})}\BibitemShut {NoStop}%
\bibitem [{\citenamefont {Grover}(1996)}]{grover}%
  \BibitemOpen
  \bibfield  {author} {\bibinfo {author} {\bibfnamefont {L.~K.}\ \bibnamefont {Grover}},\ }\bibfield  {title} {\bibinfo {title} {A fast quantum mechanical algorithm for database search},\ }in\ \href@noop {} {\emph {\bibinfo {booktitle} {Proceedings of the Twenty-Eighth Annual ACM Symposium on Theory of Computing}}},\ \bibinfo {series and number} {STOC '96}\ (\bibinfo  {publisher} {Association for Computing Machinery},\ \bibinfo {address} {New York, NY, USA},\ \bibinfo {year} {1996})\ p.\ \bibinfo {pages} {212–219}\BibitemShut {NoStop}%
\bibitem [{\citenamefont {Yang}\ \emph {et~al.}(2009)\citenamefont {Yang}, \citenamefont {Wei}, \citenamefont {Zhou}, \citenamefont {Chang},\ and\ \citenamefont {Feng}}]{yang2009solution}%
  \BibitemOpen
  \bibfield  {author} {\bibinfo {author} {\bibfnamefont {W.-L.}\ \bibnamefont {Yang}}, \bibinfo {author} {\bibfnamefont {H.}~\bibnamefont {Wei}}, \bibinfo {author} {\bibfnamefont {F.}~\bibnamefont {Zhou}}, \bibinfo {author} {\bibfnamefont {W.-L.}\ \bibnamefont {Chang}},\ and\ \bibinfo {author} {\bibfnamefont {M.}~\bibnamefont {Feng}},\ }\bibfield  {title} {\bibinfo {title} {Solution to the satisfiability problem using a complete {G}rover search with trapped ions},\ }\href@noop {} {\bibfield  {journal} {\bibinfo  {journal} {J. Phys. B: At. Mol. Opt. Phys.}\ }\textbf {\bibinfo {volume} {42}},\ \bibinfo {pages} {145503} (\bibinfo {year} {2009})}\BibitemShut {NoStop}%
\bibitem [{\citenamefont {Sundar}\ \emph {et~al.}(2019)\citenamefont {Sundar}, \citenamefont {Paredes}, \citenamefont {Damanik}, \citenamefont {Duenas-Osorio},\ and\ \citenamefont {Hazzard}}]{sundar2019quantum}%
  \BibitemOpen
  \bibfield  {author} {\bibinfo {author} {\bibfnamefont {B.}~\bibnamefont {Sundar}}, \bibinfo {author} {\bibfnamefont {R.}~\bibnamefont {Paredes}}, \bibinfo {author} {\bibfnamefont {D.~T.}\ \bibnamefont {Damanik}}, \bibinfo {author} {\bibfnamefont {L.}~\bibnamefont {Duenas-Osorio}},\ and\ \bibinfo {author} {\bibfnamefont {K.~R.}\ \bibnamefont {Hazzard}},\ }\bibfield  {title} {\bibinfo {title} {A quantum algorithm to count weighted ground states of classical spin {H}amiltonians},\ }\href@noop {} {\bibfield  {journal} {\bibinfo  {journal} {arXiv preprint arXiv:1908.01745}\ } (\bibinfo {year} {2019})}\BibitemShut {NoStop}%
\bibitem [{\citenamefont {Bartschi}\ and\ \citenamefont {Eidenbenz}(2020)}]{Bartschi2020GroverPreparation}%
  \BibitemOpen
  \bibfield  {author} {\bibinfo {author} {\bibfnamefont {A.}~\bibnamefont {Bartschi}}\ and\ \bibinfo {author} {\bibfnamefont {S.}~\bibnamefont {Eidenbenz}},\ }\bibfield  {title} {\bibinfo {title} {{G}rover mixers for {QAOA}: Shifting complexity from mixer design to state preparation},\ }\href@noop {} {\bibfield  {journal} {\bibinfo  {journal} {Proceedings - IEEE International Conference on Quantum Computing and Engineering, QCE 2020}\ ,\ \bibinfo {pages} {72}} (\bibinfo {year} {2020})}\BibitemShut {NoStop}%
\bibitem [{\citenamefont {Golden}\ \emph {et~al.}(2023)\citenamefont {Golden}, \citenamefont {B{\"a}rtschi}, \citenamefont {O'Malley},\ and\ \citenamefont {Eidenbenz}}]{golden2023quantum}%
  \BibitemOpen
  \bibfield  {author} {\bibinfo {author} {\bibfnamefont {J.}~\bibnamefont {Golden}}, \bibinfo {author} {\bibfnamefont {A.}~\bibnamefont {B{\"a}rtschi}}, \bibinfo {author} {\bibfnamefont {D.}~\bibnamefont {O'Malley}},\ and\ \bibinfo {author} {\bibfnamefont {S.}~\bibnamefont {Eidenbenz}},\ }\bibfield  {title} {\bibinfo {title} {The quantum alternating operator ansatz for satisfiability problems},\ }\href@noop {} {\bibfield  {journal} {\bibinfo  {journal} {arXiv preprint arXiv:2301.11292}\ } (\bibinfo {year} {2023})}\BibitemShut {NoStop}%
\bibitem [{\citenamefont {Mandl}\ \emph {et~al.}(2023)\citenamefont {Mandl}, \citenamefont {Barzen}, \citenamefont {Bechtold}, \citenamefont {Leymann},\ and\ \citenamefont {Wild}}]{mandl2023amplitude}%
  \BibitemOpen
  \bibfield  {author} {\bibinfo {author} {\bibfnamefont {A.}~\bibnamefont {Mandl}}, \bibinfo {author} {\bibfnamefont {J.}~\bibnamefont {Barzen}}, \bibinfo {author} {\bibfnamefont {M.}~\bibnamefont {Bechtold}}, \bibinfo {author} {\bibfnamefont {F.}~\bibnamefont {Leymann}},\ and\ \bibinfo {author} {\bibfnamefont {K.}~\bibnamefont {Wild}},\ }\bibfield  {title} {\bibinfo {title} {Amplitude amplification-inspired {QAOA}: Improving the success probability for solving 3{SAT}},\ }\href@noop {} {\bibfield  {journal} {\bibinfo  {journal} {arXiv preprint arXiv:2303.01183}\ } (\bibinfo {year} {2023})}\BibitemShut {NoStop}%
\bibitem [{\citenamefont {Kirkpatrick}\ and\ \citenamefont {Selman}(1994)}]{kirkpatrick1994critical}%
  \BibitemOpen
  \bibfield  {author} {\bibinfo {author} {\bibfnamefont {S.}~\bibnamefont {Kirkpatrick}}\ and\ \bibinfo {author} {\bibfnamefont {B.}~\bibnamefont {Selman}},\ }\bibfield  {title} {\bibinfo {title} {Critical behavior in the satisfiability of random boolean expressions},\ }\href@noop {} {\bibfield  {journal} {\bibinfo  {journal} {Science}\ }\textbf {\bibinfo {volume} {264}},\ \bibinfo {pages} {1297} (\bibinfo {year} {1994})}\BibitemShut {NoStop}%
\bibitem [{\citenamefont {M{\'e}zard}\ \emph {et~al.}(2002)\citenamefont {M{\'e}zard}, \citenamefont {Parisi},\ and\ \citenamefont {Zecchina}}]{mezard2002analytic}%
  \BibitemOpen
  \bibfield  {author} {\bibinfo {author} {\bibfnamefont {M.}~\bibnamefont {M{\'e}zard}}, \bibinfo {author} {\bibfnamefont {G.}~\bibnamefont {Parisi}},\ and\ \bibinfo {author} {\bibfnamefont {R.}~\bibnamefont {Zecchina}},\ }\bibfield  {title} {\bibinfo {title} {Analytic and algorithmic solution of random satisfiability problems},\ }\href@noop {} {\bibfield  {journal} {\bibinfo  {journal} {Science}\ }\textbf {\bibinfo {volume} {297}},\ \bibinfo {pages} {812} (\bibinfo {year} {2002})}\BibitemShut {NoStop}%
\bibitem [{\citenamefont {Asano}\ and\ \citenamefont {Williamson}(2002)}]{asano2002improved}%
  \BibitemOpen
  \bibfield  {author} {\bibinfo {author} {\bibfnamefont {T.}~\bibnamefont {Asano}}\ and\ \bibinfo {author} {\bibfnamefont {D.~P.}\ \bibnamefont {Williamson}},\ }\bibfield  {title} {\bibinfo {title} {Improved approximation algorithms for {MAX SAT}},\ }\href@noop {} {\bibfield  {journal} {\bibinfo  {journal} {Journal of Algorithms}\ }\textbf {\bibinfo {volume} {42}},\ \bibinfo {pages} {173} (\bibinfo {year} {2002})}\BibitemShut {NoStop}%
\bibitem [{\citenamefont {Nannicini}(2019)}]{nannicini2019performance}%
  \BibitemOpen
  \bibfield  {author} {\bibinfo {author} {\bibfnamefont {G.}~\bibnamefont {Nannicini}},\ }\bibfield  {title} {\bibinfo {title} {Performance of hybrid quantum-classical variational heuristics for combinatorial optimization},\ }\href@noop {} {\bibfield  {journal} {\bibinfo  {journal} {Phys. Rev. E}\ }\textbf {\bibinfo {volume} {99}},\ \bibinfo {pages} {013304} (\bibinfo {year} {2019})}\BibitemShut {NoStop}%
\bibitem [{\citenamefont {M{\'e}zard}\ and\ \citenamefont {Zecchina}(2002)}]{Mezard2002RandomAlgorithm}%
  \BibitemOpen
  \bibfield  {author} {\bibinfo {author} {\bibfnamefont {M.}~\bibnamefont {M{\'e}zard}}\ and\ \bibinfo {author} {\bibfnamefont {R.}~\bibnamefont {Zecchina}},\ }\bibfield  {title} {\bibinfo {title} {Random k-satisfiability problem: From an analytic solution to an efficient algorithm},\ }\href@noop {} {\bibfield  {journal} {\bibinfo  {journal} {Phys. Rev. E}\ }\textbf {\bibinfo {volume} {66}},\ \bibinfo {pages} {056126} (\bibinfo {year} {2002})}\BibitemShut {NoStop}%
\bibitem [{\citenamefont {Chen}\ \emph {et~al.}(2009)\citenamefont {Chen}, \citenamefont {Safarpour}, \citenamefont {Veneris},\ and\ \citenamefont {Marques-Silva}}]{chen2009spatial}%
  \BibitemOpen
  \bibfield  {author} {\bibinfo {author} {\bibfnamefont {Y.}~\bibnamefont {Chen}}, \bibinfo {author} {\bibfnamefont {S.}~\bibnamefont {Safarpour}}, \bibinfo {author} {\bibfnamefont {A.}~\bibnamefont {Veneris}},\ and\ \bibinfo {author} {\bibfnamefont {J.}~\bibnamefont {Marques-Silva}},\ }\bibfield  {title} {\bibinfo {title} {Spatial and temporal design debug using partial {MaxSAT}},\ }in\ \href@noop {} {\emph {\bibinfo {booktitle} {Proceedings of the 19th ACM Great Lakes symposium on VLSI}}}\ (\bibinfo {year} {2009})\ pp.\ \bibinfo {pages} {345--350}\BibitemShut {NoStop}%
\bibitem [{\citenamefont {Dimitrova}\ \emph {et~al.}(2018)\citenamefont {Dimitrova}, \citenamefont {Ghasemi},\ and\ \citenamefont {Topcu}}]{dimitrova2018maximum}%
  \BibitemOpen
  \bibfield  {author} {\bibinfo {author} {\bibfnamefont {R.}~\bibnamefont {Dimitrova}}, \bibinfo {author} {\bibfnamefont {M.}~\bibnamefont {Ghasemi}},\ and\ \bibinfo {author} {\bibfnamefont {U.}~\bibnamefont {Topcu}},\ }\bibfield  {title} {\bibinfo {title} {Maximum realizability for linear temporal logic specifications},\ }in\ \href@noop {} {\emph {\bibinfo {booktitle} {Automated Technology for Verification and Analysis: 16th International Symposium, ATVA 2018, Los Angeles, CA, USA, October 7-10, 2018, Proceedings 16}}}\ (\bibinfo {organization} {Springer},\ \bibinfo {year} {2018})\ pp.\ \bibinfo {pages} {458--475}\BibitemShut {NoStop}%
\bibitem [{\citenamefont {Malioutov}\ and\ \citenamefont {Meel}(2018)}]{malioutov2018mlic}%
  \BibitemOpen
  \bibfield  {author} {\bibinfo {author} {\bibfnamefont {D.}~\bibnamefont {Malioutov}}\ and\ \bibinfo {author} {\bibfnamefont {K.~S.}\ \bibnamefont {Meel}},\ }\bibfield  {title} {\bibinfo {title} {Mlic: A {MaxSAT}-based framework for learning interpretable classification rules},\ }in\ \href@noop {} {\emph {\bibinfo {booktitle} {International Conference on Principles and Practice of Constraint Programming}}}\ (\bibinfo {organization} {Springer},\ \bibinfo {year} {2018})\ pp.\ \bibinfo {pages} {312--327}\BibitemShut {NoStop}%
\bibitem [{\citenamefont {Berg}\ \emph {et~al.}(2019)\citenamefont {Berg}, \citenamefont {Hyttinen},\ and\ \citenamefont {J{\"a}rvisalo}}]{berg2019applications}%
  \BibitemOpen
  \bibfield  {author} {\bibinfo {author} {\bibfnamefont {O.~J.}\ \bibnamefont {Berg}}, \bibinfo {author} {\bibfnamefont {A.~J.}\ \bibnamefont {Hyttinen}},\ and\ \bibinfo {author} {\bibfnamefont {M.~J.}\ \bibnamefont {J{\"a}rvisalo}},\ }\bibfield  {title} {\bibinfo {title} {Applications of {MaxSAT} in data analysis},\ }\href@noop {} {\bibfield  {journal} {\bibinfo  {journal} {Proceedings of Pragmatics of SAT 2015 and 2018}\ } (\bibinfo {year} {2019})}\BibitemShut {NoStop}%
\bibitem [{\citenamefont {Guerra}\ and\ \citenamefont {Lynce}(2012)}]{guerra2012reasoning}%
  \BibitemOpen
  \bibfield  {author} {\bibinfo {author} {\bibfnamefont {J.}~\bibnamefont {Guerra}}\ and\ \bibinfo {author} {\bibfnamefont {I.}~\bibnamefont {Lynce}},\ }\bibfield  {title} {\bibinfo {title} {Reasoning over biological networks using maximum satisfiability},\ }in\ \href@noop {} {\emph {\bibinfo {booktitle} {International conference on principles and practice of constraint programming}}}\ (\bibinfo {organization} {Springer},\ \bibinfo {year} {2012})\ pp.\ \bibinfo {pages} {941--956}\BibitemShut {NoStop}%
\bibitem [{\citenamefont {Martins}(2017)}]{martins2017solving}%
  \BibitemOpen
  \bibfield  {author} {\bibinfo {author} {\bibfnamefont {R.}~\bibnamefont {Martins}},\ }\bibfield  {title} {\bibinfo {title} {Solving {RNA} alignment with {MaxSAT}},\ }\href@noop {} {\bibfield  {journal} {\bibinfo  {journal} {MaxSAT Evaluation 2017}\ ,\ \bibinfo {pages} {29}} (\bibinfo {year} {2017})}\BibitemShut {NoStop}%
\bibitem [{\citenamefont {Matsuda}\ \emph {et~al.}(2009)\citenamefont {Matsuda}, \citenamefont {Nishimori},\ and\ \citenamefont {Katzgraber}}]{matsuda2009ground}%
  \BibitemOpen
  \bibfield  {author} {\bibinfo {author} {\bibfnamefont {Y.}~\bibnamefont {Matsuda}}, \bibinfo {author} {\bibfnamefont {H.}~\bibnamefont {Nishimori}},\ and\ \bibinfo {author} {\bibfnamefont {H.~G.}\ \bibnamefont {Katzgraber}},\ }\bibfield  {title} {\bibinfo {title} {Ground-state statistics from annealing algorithms: quantum versus classical approaches},\ }\href@noop {} {\bibfield  {journal} {\bibinfo  {journal} {New J. Phys.}\ }\textbf {\bibinfo {volume} {11}},\ \bibinfo {pages} {073021} (\bibinfo {year} {2009})}\BibitemShut {NoStop}%
\bibitem [{\citenamefont {Bittel}\ and\ \citenamefont {Kliesch}(2021)}]{bittel2021training}%
  \BibitemOpen
  \bibfield  {author} {\bibinfo {author} {\bibfnamefont {L.}~\bibnamefont {Bittel}}\ and\ \bibinfo {author} {\bibfnamefont {M.}~\bibnamefont {Kliesch}},\ }\bibfield  {title} {\bibinfo {title} {Training variational quantum algorithms is {NP}-hard},\ }\href@noop {} {\bibfield  {journal} {\bibinfo  {journal} {Phys. Rev. Lett.}\ }\textbf {\bibinfo {volume} {127}},\ \bibinfo {pages} {120502} (\bibinfo {year} {2021})}\BibitemShut {NoStop}%
\bibitem [{\citenamefont {Shaydulin}\ \emph {et~al.}(2019)\citenamefont {Shaydulin}, \citenamefont {Safro},\ and\ \citenamefont {Larson}}]{shaydulin2019multistart}%
  \BibitemOpen
  \bibfield  {author} {\bibinfo {author} {\bibfnamefont {R.}~\bibnamefont {Shaydulin}}, \bibinfo {author} {\bibfnamefont {I.}~\bibnamefont {Safro}},\ and\ \bibinfo {author} {\bibfnamefont {J.}~\bibnamefont {Larson}},\ }\bibfield  {title} {\bibinfo {title} {Multistart methods for quantum approximate optimization},\ }in\ \href@noop {} {\emph {\bibinfo {booktitle} {2019 IEEE high performance extreme computing conference (HPEC)}}}\ (\bibinfo {organization} {IEEE},\ \bibinfo {year} {2019})\ pp.\ \bibinfo {pages} {1--8}\BibitemShut {NoStop}%
\bibitem [{\citenamefont {McClean}\ \emph {et~al.}(2018)\citenamefont {McClean}, \citenamefont {Boixo}, \citenamefont {Smelyanskiy}, \citenamefont {Babbush},\ and\ \citenamefont {Neven}}]{mcclean2018barren}%
  \BibitemOpen
  \bibfield  {author} {\bibinfo {author} {\bibfnamefont {J.~R.}\ \bibnamefont {McClean}}, \bibinfo {author} {\bibfnamefont {S.}~\bibnamefont {Boixo}}, \bibinfo {author} {\bibfnamefont {V.~N.}\ \bibnamefont {Smelyanskiy}}, \bibinfo {author} {\bibfnamefont {R.}~\bibnamefont {Babbush}},\ and\ \bibinfo {author} {\bibfnamefont {H.}~\bibnamefont {Neven}},\ }\bibfield  {title} {\bibinfo {title} {Barren plateaus in quantum neural network training landscapes},\ }\href@noop {} {\bibfield  {journal} {\bibinfo  {journal} {Nat. Commun.}\ }\textbf {\bibinfo {volume} {9}},\ \bibinfo {pages} {4812} (\bibinfo {year} {2018})}\BibitemShut {NoStop}%
\bibitem [{\citenamefont {S\o{}rensen}\ and\ \citenamefont {M\o{}lmer}(1999)}]{sorensen:quantum_1999}%
  \BibitemOpen
  \bibfield  {author} {\bibinfo {author} {\bibfnamefont {A.}~\bibnamefont {S\o{}rensen}}\ and\ \bibinfo {author} {\bibfnamefont {K.}~\bibnamefont {M\o{}lmer}},\ }\bibfield  {title} {\bibinfo {title} {Quantum computation with ions in thermal motion},\ }\href@noop {} {\bibfield  {journal} {\bibinfo  {journal} {Phys. Rev. Lett.}\ }\textbf {\bibinfo {volume} {82}},\ \bibinfo {pages} {1971} (\bibinfo {year} {1999})}\BibitemShut {NoStop}%
\bibitem [{Note1()}]{Note1}%
  \BibitemOpen
  \bibinfo {note} {4-qubit example: \{[-0, +2, +3], [+0, +2, -3], [-1, +2, -3], [-1, -2, -3], [-1, -2, +3], [+1, +2, -3], [+0, +2, +3], [-0, +1, -3]\}; 5-qubit example: \{[+1, -2, -3], [-1, -3, +4], [+0, -2, -4], [-0, -2, +3], [+2, +3, +4], [-0, +1, +2], [+0, -2, 4], [-1, +2, -4]\}. This notation lists the set of conjunctive clauses, where the notation for each clause is that minus signs imply the negation of a Boolean variable, and the number indexes the Boolean variable. For example $[-0, +2, +3]=(\protect \bar {x}_0\vee x_2 \vee x_3)$.}\BibitemShut {Stop}%
\bibitem [{\citenamefont {Golden}\ \emph {et~al.}(2022)\citenamefont {Golden}, \citenamefont {B{\"a}rtschi}, \citenamefont {O’Malley},\ and\ \citenamefont {Eidenbenz}}]{golden2022fair}%
  \BibitemOpen
  \bibfield  {author} {\bibinfo {author} {\bibfnamefont {J.}~\bibnamefont {Golden}}, \bibinfo {author} {\bibfnamefont {A.}~\bibnamefont {B{\"a}rtschi}}, \bibinfo {author} {\bibfnamefont {D.}~\bibnamefont {O’Malley}},\ and\ \bibinfo {author} {\bibfnamefont {S.}~\bibnamefont {Eidenbenz}},\ }\bibfield  {title} {\bibinfo {title} {Fair sampling error analysis on {NISQ} devices},\ }\href@noop {} {\bibfield  {journal} {\bibinfo  {journal} {ACM Transactions on Quantum Computing}\ }\textbf {\bibinfo {volume} {3}},\ \bibinfo {pages} {1} (\bibinfo {year} {2022})}\BibitemShut {NoStop}%
\bibitem [{\citenamefont {Moses}\ \emph {et~al.}(2023{\natexlab{b}})\citenamefont {Moses}, \citenamefont {Baldwin}, \citenamefont {Allman}, \citenamefont {Ancona}, \citenamefont {Ascarrunz}, \citenamefont {Barnes}, \citenamefont {Bartolotta}, \citenamefont {Bjork}, \citenamefont {Blanchard}, \citenamefont {Bohn}, \citenamefont {Bohnet}, \citenamefont {Brown}, \citenamefont {Burdick}, \citenamefont {Burton}, \citenamefont {Campbell}, \citenamefont {Campora}, \citenamefont {Carron}, \citenamefont {Chambers}, \citenamefont {Chan}, \citenamefont {Chen}, \citenamefont {Chernoguzov}, \citenamefont {Chertkov}, \citenamefont {Colina}, \citenamefont {Curtis}, \citenamefont {Daniel}, \citenamefont {DeCross}, \citenamefont {Deen}, \citenamefont {Delaney}, \citenamefont {Dreiling}, \citenamefont {Ertsgaard}, \citenamefont {Esposito}, \citenamefont {Estey}, \citenamefont {Fabrikant}, \citenamefont {Figgatt}, \citenamefont {Foltz}, \citenamefont {Foss-Feig}, \citenamefont {Francois}, \citenamefont {Gaebler}, \citenamefont
  {Gatterman}, \citenamefont {Gilbreth}, \citenamefont {Giles}, \citenamefont {Glynn}, \citenamefont {Hall}, \citenamefont {Hankin}, \citenamefont {Hansen}, \citenamefont {Hayes}, \citenamefont {Higashi}, \citenamefont {Hoffman}, \citenamefont {Horning}, \citenamefont {Hout}, \citenamefont {Jacobs}, \citenamefont {Johansen}, \citenamefont {Jones}, \citenamefont {Karcz}, \citenamefont {Klein}, \citenamefont {Lauria}, \citenamefont {Lee}, \citenamefont {Liefer}, \citenamefont {Lu}, \citenamefont {Lucchetti}, \citenamefont {Lytle}, \citenamefont {Malm}, \citenamefont {Matheny}, \citenamefont {Mathewson}, \citenamefont {Mayer}, \citenamefont {Miller}, \citenamefont {Mills}, \citenamefont {Neyenhuis}, \citenamefont {Nugent}, \citenamefont {Olson}, \citenamefont {Parks}, \citenamefont {Price}, \citenamefont {Price}, \citenamefont {Pugh}, \citenamefont {Ransford}, \citenamefont {Reed}, \citenamefont {Roman}, \citenamefont {Rowe}, \citenamefont {Ryan-Anderson}, \citenamefont {Sanders}, \citenamefont {Sedlacek},
  \citenamefont {Shevchuk}, \citenamefont {Siegfried}, \citenamefont {Skripka}, \citenamefont {Spaun}, \citenamefont {Sprenkle}, \citenamefont {Stutz}, \citenamefont {Swallows}, \citenamefont {Tobey}, \citenamefont {Tran}, \citenamefont {Tran}, \citenamefont {Vogt}, \citenamefont {Volin}, \citenamefont {Walker}, \citenamefont {Zolot},\ and\ \citenamefont {Pino}}]{moses:race-track_2023}%
  \BibitemOpen
  \bibfield  {author} {\bibinfo {author} {\bibfnamefont {S.~A.}\ \bibnamefont {Moses}}, \bibinfo {author} {\bibfnamefont {C.~H.}\ \bibnamefont {Baldwin}}, \bibinfo {author} {\bibfnamefont {M.~S.}\ \bibnamefont {Allman}}, \bibinfo {author} {\bibfnamefont {R.}~\bibnamefont {Ancona}}, \bibinfo {author} {\bibfnamefont {L.}~\bibnamefont {Ascarrunz}}, \bibinfo {author} {\bibfnamefont {C.}~\bibnamefont {Barnes}}, \bibinfo {author} {\bibfnamefont {J.}~\bibnamefont {Bartolotta}}, \bibinfo {author} {\bibfnamefont {B.}~\bibnamefont {Bjork}}, \bibinfo {author} {\bibfnamefont {P.}~\bibnamefont {Blanchard}}, \bibinfo {author} {\bibfnamefont {M.}~\bibnamefont {Bohn}}, \bibinfo {author} {\bibfnamefont {J.~G.}\ \bibnamefont {Bohnet}}, \bibinfo {author} {\bibfnamefont {N.~C.}\ \bibnamefont {Brown}}, \bibinfo {author} {\bibfnamefont {N.~Q.}\ \bibnamefont {Burdick}}, \bibinfo {author} {\bibfnamefont {W.~C.}\ \bibnamefont {Burton}}, \bibinfo {author} {\bibfnamefont {S.~L.}\ \bibnamefont {Campbell}}, \bibinfo {author}
  {\bibfnamefont {J.~P.}\ \bibnamefont {Campora}}, \bibinfo {author} {\bibfnamefont {C.}~\bibnamefont {Carron}}, \bibinfo {author} {\bibfnamefont {J.}~\bibnamefont {Chambers}}, \bibinfo {author} {\bibfnamefont {J.~W.}\ \bibnamefont {Chan}}, \bibinfo {author} {\bibfnamefont {Y.~H.}\ \bibnamefont {Chen}}, \bibinfo {author} {\bibfnamefont {A.}~\bibnamefont {Chernoguzov}}, \bibinfo {author} {\bibfnamefont {E.}~\bibnamefont {Chertkov}}, \bibinfo {author} {\bibfnamefont {J.}~\bibnamefont {Colina}}, \bibinfo {author} {\bibfnamefont {J.~P.}\ \bibnamefont {Curtis}}, \bibinfo {author} {\bibfnamefont {R.}~\bibnamefont {Daniel}}, \bibinfo {author} {\bibfnamefont {M.}~\bibnamefont {DeCross}}, \bibinfo {author} {\bibfnamefont {D.}~\bibnamefont {Deen}}, \bibinfo {author} {\bibfnamefont {C.}~\bibnamefont {Delaney}}, \bibinfo {author} {\bibfnamefont {J.~M.}\ \bibnamefont {Dreiling}}, \bibinfo {author} {\bibfnamefont {C.~T.}\ \bibnamefont {Ertsgaard}}, \bibinfo {author} {\bibfnamefont {J.}~\bibnamefont {Esposito}}, \bibinfo
  {author} {\bibfnamefont {B.}~\bibnamefont {Estey}}, \bibinfo {author} {\bibfnamefont {M.}~\bibnamefont {Fabrikant}}, \bibinfo {author} {\bibfnamefont {C.}~\bibnamefont {Figgatt}}, \bibinfo {author} {\bibfnamefont {C.}~\bibnamefont {Foltz}}, \bibinfo {author} {\bibfnamefont {M.}~\bibnamefont {Foss-Feig}}, \bibinfo {author} {\bibfnamefont {D.}~\bibnamefont {Francois}}, \bibinfo {author} {\bibfnamefont {J.~P.}\ \bibnamefont {Gaebler}}, \bibinfo {author} {\bibfnamefont {T.~M.}\ \bibnamefont {Gatterman}}, \bibinfo {author} {\bibfnamefont {C.~N.}\ \bibnamefont {Gilbreth}}, \bibinfo {author} {\bibfnamefont {J.}~\bibnamefont {Giles}}, \bibinfo {author} {\bibfnamefont {E.}~\bibnamefont {Glynn}}, \bibinfo {author} {\bibfnamefont {A.}~\bibnamefont {Hall}}, \bibinfo {author} {\bibfnamefont {A.~M.}\ \bibnamefont {Hankin}}, \bibinfo {author} {\bibfnamefont {A.}~\bibnamefont {Hansen}}, \bibinfo {author} {\bibfnamefont {D.}~\bibnamefont {Hayes}}, \bibinfo {author} {\bibfnamefont {B.}~\bibnamefont {Higashi}}, \bibinfo
  {author} {\bibfnamefont {I.~M.}\ \bibnamefont {Hoffman}}, \bibinfo {author} {\bibfnamefont {B.}~\bibnamefont {Horning}}, \bibinfo {author} {\bibfnamefont {J.~J.}\ \bibnamefont {Hout}}, \bibinfo {author} {\bibfnamefont {R.}~\bibnamefont {Jacobs}}, \bibinfo {author} {\bibfnamefont {J.}~\bibnamefont {Johansen}}, \bibinfo {author} {\bibfnamefont {L.}~\bibnamefont {Jones}}, \bibinfo {author} {\bibfnamefont {J.}~\bibnamefont {Karcz}}, \bibinfo {author} {\bibfnamefont {T.}~\bibnamefont {Klein}}, \bibinfo {author} {\bibfnamefont {P.}~\bibnamefont {Lauria}}, \bibinfo {author} {\bibfnamefont {P.}~\bibnamefont {Lee}}, \bibinfo {author} {\bibfnamefont {D.}~\bibnamefont {Liefer}}, \bibinfo {author} {\bibfnamefont {S.~T.}\ \bibnamefont {Lu}}, \bibinfo {author} {\bibfnamefont {D.}~\bibnamefont {Lucchetti}}, \bibinfo {author} {\bibfnamefont {C.}~\bibnamefont {Lytle}}, \bibinfo {author} {\bibfnamefont {A.}~\bibnamefont {Malm}}, \bibinfo {author} {\bibfnamefont {M.}~\bibnamefont {Matheny}}, \bibinfo {author} {\bibfnamefont
  {B.}~\bibnamefont {Mathewson}}, \bibinfo {author} {\bibfnamefont {K.}~\bibnamefont {Mayer}}, \bibinfo {author} {\bibfnamefont {D.~B.}\ \bibnamefont {Miller}}, \bibinfo {author} {\bibfnamefont {M.}~\bibnamefont {Mills}}, \bibinfo {author} {\bibfnamefont {B.}~\bibnamefont {Neyenhuis}}, \bibinfo {author} {\bibfnamefont {L.}~\bibnamefont {Nugent}}, \bibinfo {author} {\bibfnamefont {S.}~\bibnamefont {Olson}}, \bibinfo {author} {\bibfnamefont {J.}~\bibnamefont {Parks}}, \bibinfo {author} {\bibfnamefont {G.~N.}\ \bibnamefont {Price}}, \bibinfo {author} {\bibfnamefont {Z.}~\bibnamefont {Price}}, \bibinfo {author} {\bibfnamefont {M.}~\bibnamefont {Pugh}}, \bibinfo {author} {\bibfnamefont {A.}~\bibnamefont {Ransford}}, \bibinfo {author} {\bibfnamefont {A.~P.}\ \bibnamefont {Reed}}, \bibinfo {author} {\bibfnamefont {C.}~\bibnamefont {Roman}}, \bibinfo {author} {\bibfnamefont {M.}~\bibnamefont {Rowe}}, \bibinfo {author} {\bibfnamefont {C.}~\bibnamefont {Ryan-Anderson}}, \bibinfo {author} {\bibfnamefont
  {S.}~\bibnamefont {Sanders}}, \bibinfo {author} {\bibfnamefont {J.}~\bibnamefont {Sedlacek}}, \bibinfo {author} {\bibfnamefont {P.}~\bibnamefont {Shevchuk}}, \bibinfo {author} {\bibfnamefont {P.}~\bibnamefont {Siegfried}}, \bibinfo {author} {\bibfnamefont {T.}~\bibnamefont {Skripka}}, \bibinfo {author} {\bibfnamefont {B.}~\bibnamefont {Spaun}}, \bibinfo {author} {\bibfnamefont {R.~T.}\ \bibnamefont {Sprenkle}}, \bibinfo {author} {\bibfnamefont {R.~P.}\ \bibnamefont {Stutz}}, \bibinfo {author} {\bibfnamefont {M.}~\bibnamefont {Swallows}}, \bibinfo {author} {\bibfnamefont {R.~I.}\ \bibnamefont {Tobey}}, \bibinfo {author} {\bibfnamefont {A.}~\bibnamefont {Tran}}, \bibinfo {author} {\bibfnamefont {T.}~\bibnamefont {Tran}}, \bibinfo {author} {\bibfnamefont {E.}~\bibnamefont {Vogt}}, \bibinfo {author} {\bibfnamefont {C.}~\bibnamefont {Volin}}, \bibinfo {author} {\bibfnamefont {J.}~\bibnamefont {Walker}}, \bibinfo {author} {\bibfnamefont {A.~M.}\ \bibnamefont {Zolot}},\ and\ \bibinfo {author} {\bibfnamefont
  {J.~M.}\ \bibnamefont {Pino}},\ }\bibfield  {title} {\bibinfo {title} {A race-track trapped-ion quantum processor},\ }\href@noop {} {\bibfield  {journal} {\bibinfo  {journal} {Phys. Rev. X}\ }\textbf {\bibinfo {volume} {13}},\ \bibinfo {pages} {041052} (\bibinfo {year} {2023}{\natexlab{b}})}\BibitemShut {NoStop}%
\bibitem [{\citenamefont {Dechter}(1999)}]{DECHTER199941}%
  \BibitemOpen
  \bibfield  {author} {\bibinfo {author} {\bibfnamefont {R.}~\bibnamefont {Dechter}},\ }\bibfield  {title} {\bibinfo {title} {Bucket elimination: A unifying framework for reasoning},\ }\href {https://doi.org/https://doi.org/10.1016/S0004-3702(99)00059-4} {\bibfield  {journal} {\bibinfo  {journal} {Artificial Intelligence}\ }\textbf {\bibinfo {volume} {113}},\ \bibinfo {pages} {41} (\bibinfo {year} {1999})}\BibitemShut {NoStop}%
\bibitem [{\citenamefont {Akshay}\ \emph {et~al.}(2021)\citenamefont {Akshay}, \citenamefont {Rabinovich}, \citenamefont {Campos},\ and\ \citenamefont {Biamonte}}]{akshay2021parameter}%
  \BibitemOpen
  \bibfield  {author} {\bibinfo {author} {\bibfnamefont {V.}~\bibnamefont {Akshay}}, \bibinfo {author} {\bibfnamefont {D.}~\bibnamefont {Rabinovich}}, \bibinfo {author} {\bibfnamefont {E.}~\bibnamefont {Campos}},\ and\ \bibinfo {author} {\bibfnamefont {J.}~\bibnamefont {Biamonte}},\ }\bibfield  {title} {\bibinfo {title} {Parameter concentrations in quantum approximate optimization},\ }\href@noop {} {\bibfield  {journal} {\bibinfo  {journal} {Phys. Rev. A}\ }\textbf {\bibinfo {volume} {104}},\ \bibinfo {pages} {L010401} (\bibinfo {year} {2021})}\BibitemShut {NoStop}%
\bibitem [{\citenamefont {Wang}\ \emph {et~al.}(2021)\citenamefont {Wang}, \citenamefont {Fontana}, \citenamefont {Cerezo}, \citenamefont {Sharma}, \citenamefont {Sone}, \citenamefont {Cincio},\ and\ \citenamefont {Coles}}]{wang2021noise}%
  \BibitemOpen
  \bibfield  {author} {\bibinfo {author} {\bibfnamefont {S.}~\bibnamefont {Wang}}, \bibinfo {author} {\bibfnamefont {E.}~\bibnamefont {Fontana}}, \bibinfo {author} {\bibfnamefont {M.}~\bibnamefont {Cerezo}}, \bibinfo {author} {\bibfnamefont {K.}~\bibnamefont {Sharma}}, \bibinfo {author} {\bibfnamefont {A.}~\bibnamefont {Sone}}, \bibinfo {author} {\bibfnamefont {L.}~\bibnamefont {Cincio}},\ and\ \bibinfo {author} {\bibfnamefont {P.~J.}\ \bibnamefont {Coles}},\ }\bibfield  {title} {\bibinfo {title} {Noise-induced barren plateaus in variational quantum algorithms},\ }\href@noop {} {\bibfield  {journal} {\bibinfo  {journal} {Nat. Commun.}\ }\textbf {\bibinfo {volume} {12}},\ \bibinfo {pages} {6961} (\bibinfo {year} {2021})}\BibitemShut {NoStop}%
\bibitem [{\citenamefont {Babbush}\ \emph {et~al.}(2021)\citenamefont {Babbush}, \citenamefont {McClean}, \citenamefont {Newman}, \citenamefont {Gidney}, \citenamefont {Boixo},\ and\ \citenamefont {Neven}}]{babbush2021}%
  \BibitemOpen
  \bibfield  {author} {\bibinfo {author} {\bibfnamefont {R.}~\bibnamefont {Babbush}}, \bibinfo {author} {\bibfnamefont {J.~R.}\ \bibnamefont {McClean}}, \bibinfo {author} {\bibfnamefont {M.}~\bibnamefont {Newman}}, \bibinfo {author} {\bibfnamefont {C.}~\bibnamefont {Gidney}}, \bibinfo {author} {\bibfnamefont {S.}~\bibnamefont {Boixo}},\ and\ \bibinfo {author} {\bibfnamefont {H.}~\bibnamefont {Neven}},\ }\bibfield  {title} {\bibinfo {title} {Focus beyond quadratic speedups for error-corrected quantum advantage},\ }\href@noop {} {\bibfield  {journal} {\bibinfo  {journal} {PRX Quantum}\ }\textbf {\bibinfo {volume} {2}},\ \bibinfo {pages} {010103} (\bibinfo {year} {2021})}\BibitemShut {NoStop}%
\bibitem [{\citenamefont {Beverland}\ \emph {et~al.}(2022)\citenamefont {Beverland}, \citenamefont {Murali}, \citenamefont {Troyer}, \citenamefont {Svore}, \citenamefont {Hoefler}, \citenamefont {Kliuchnikov}, \citenamefont {Low}, \citenamefont {Soeken}, \citenamefont {Sundaram},\ and\ \citenamefont {Vaschillo}}]{beverland2022assessing}%
  \BibitemOpen
  \bibfield  {author} {\bibinfo {author} {\bibfnamefont {M.~E.}\ \bibnamefont {Beverland}}, \bibinfo {author} {\bibfnamefont {P.}~\bibnamefont {Murali}}, \bibinfo {author} {\bibfnamefont {M.}~\bibnamefont {Troyer}}, \bibinfo {author} {\bibfnamefont {K.~M.}\ \bibnamefont {Svore}}, \bibinfo {author} {\bibfnamefont {T.}~\bibnamefont {Hoefler}}, \bibinfo {author} {\bibfnamefont {V.}~\bibnamefont {Kliuchnikov}}, \bibinfo {author} {\bibfnamefont {G.~H.}\ \bibnamefont {Low}}, \bibinfo {author} {\bibfnamefont {M.}~\bibnamefont {Soeken}}, \bibinfo {author} {\bibfnamefont {A.}~\bibnamefont {Sundaram}},\ and\ \bibinfo {author} {\bibfnamefont {A.}~\bibnamefont {Vaschillo}},\ }\bibfield  {title} {\bibinfo {title} {Assessing requirements to scale to practical quantum advantage},\ }\href@noop {} {\bibfield  {journal} {\bibinfo  {journal} {arXiv preprint arXiv:2211.07629}\ } (\bibinfo {year} {2022})}\BibitemShut {NoStop}%
\bibitem [{\citenamefont {Headley}\ and\ \citenamefont {Wilhelm}(2023)}]{headley2023problem}%
  \BibitemOpen
  \bibfield  {author} {\bibinfo {author} {\bibfnamefont {D.}~\bibnamefont {Headley}}\ and\ \bibinfo {author} {\bibfnamefont {F.~K.}\ \bibnamefont {Wilhelm}},\ }\bibfield  {title} {\bibinfo {title} {Problem-size-independent angles for a {G}rover-driven quantum approximate optimization algorithm},\ }\href@noop {} {\bibfield  {journal} {\bibinfo  {journal} {Phys. Rev. A}\ }\textbf {\bibinfo {volume} {107}},\ \bibinfo {pages} {012412} (\bibinfo {year} {2023})}\BibitemShut {NoStop}%
\bibitem [{\citenamefont {Bridi}\ and\ \citenamefont {de~Lima~Marquezino}(2024)}]{bridi2024analytical}%
  \BibitemOpen
  \bibfield  {author} {\bibinfo {author} {\bibfnamefont {G.~A.}\ \bibnamefont {Bridi}}\ and\ \bibinfo {author} {\bibfnamefont {F.}~\bibnamefont {de~Lima~Marquezino}},\ }\bibfield  {title} {\bibinfo {title} {Analytical results for the quantum alternating operator ansatz with {G}rover mixer},\ }\href@noop {} {\bibfield  {journal} {\bibinfo  {journal} {arXiv preprint arXiv:2401.11056}\ } (\bibinfo {year} {2024})}\BibitemShut {NoStop}%
\bibitem [{\citenamefont {Egger}\ \emph {et~al.}(2021)\citenamefont {Egger}, \citenamefont {Mare{\v{c}}ek},\ and\ \citenamefont {Woerner}}]{egger2021warm}%
  \BibitemOpen
  \bibfield  {author} {\bibinfo {author} {\bibfnamefont {D.~J.}\ \bibnamefont {Egger}}, \bibinfo {author} {\bibfnamefont {J.}~\bibnamefont {Mare{\v{c}}ek}},\ and\ \bibinfo {author} {\bibfnamefont {S.}~\bibnamefont {Woerner}},\ }\bibfield  {title} {\bibinfo {title} {Warm-starting quantum optimization},\ }\href@noop {} {\bibfield  {journal} {\bibinfo  {journal} {Quantum}\ }\textbf {\bibinfo {volume} {5}},\ \bibinfo {pages} {479} (\bibinfo {year} {2021})}\BibitemShut {NoStop}%
\bibitem [{\citenamefont {Tate}\ \emph {et~al.}(2023)\citenamefont {Tate}, \citenamefont {Moondra}, \citenamefont {Gard}, \citenamefont {Mohler},\ and\ \citenamefont {Gupta}}]{tate2023warm}%
  \BibitemOpen
  \bibfield  {author} {\bibinfo {author} {\bibfnamefont {R.}~\bibnamefont {Tate}}, \bibinfo {author} {\bibfnamefont {J.}~\bibnamefont {Moondra}}, \bibinfo {author} {\bibfnamefont {B.}~\bibnamefont {Gard}}, \bibinfo {author} {\bibfnamefont {G.}~\bibnamefont {Mohler}},\ and\ \bibinfo {author} {\bibfnamefont {S.}~\bibnamefont {Gupta}},\ }\bibfield  {title} {\bibinfo {title} {Warm-started {QAOA} with custom mixers provably converges and computationally beats {Goemans-Williamson's Max-Cut} at low circuit depths},\ }\href@noop {} {\bibfield  {journal} {\bibinfo  {journal} {Quantum}\ }\textbf {\bibinfo {volume} {7}},\ \bibinfo {pages} {1121} (\bibinfo {year} {2023})}\BibitemShut {NoStop}%
\bibitem [{\citenamefont {Duenas-Osorio}\ \emph {et~al.}(2017)\citenamefont {Duenas-Osorio}, \citenamefont {Meel}, \citenamefont {Paredes},\ and\ \citenamefont {Vardi}}]{duenas2017counting}%
  \BibitemOpen
  \bibfield  {author} {\bibinfo {author} {\bibfnamefont {L.}~\bibnamefont {Duenas-Osorio}}, \bibinfo {author} {\bibfnamefont {K.}~\bibnamefont {Meel}}, \bibinfo {author} {\bibfnamefont {R.}~\bibnamefont {Paredes}},\ and\ \bibinfo {author} {\bibfnamefont {M.}~\bibnamefont {Vardi}},\ }\bibfield  {title} {\bibinfo {title} {Counting-based reliability estimation for power-transmission grids},\ }in\ \href@noop {} {\emph {\bibinfo {booktitle} {Proceedings of the AAAI Conference on Artificial Intelligence}}},\ Vol.~\bibinfo {volume} {31}\ (\bibinfo {year} {2017})\BibitemShut {NoStop}%
\bibitem [{\citenamefont {Maslov}(2017)}]{maslov2017basic}%
  \BibitemOpen
  \bibfield  {author} {\bibinfo {author} {\bibfnamefont {D.}~\bibnamefont {Maslov}},\ }\bibfield  {title} {\bibinfo {title} {Basic circuit compilation techniques for an ion-trap quantum machine},\ }\href@noop {} {\bibfield  {journal} {\bibinfo  {journal} {New J. Phys.}\ }\textbf {\bibinfo {volume} {19}},\ \bibinfo {pages} {023035} (\bibinfo {year} {2017})}\BibitemShut {NoStop}%
\bibitem [{\citenamefont {Li}\ and\ \citenamefont {Quan}(2010)}]{liefficient2010}%
  \BibitemOpen
  \bibfield  {author} {\bibinfo {author} {\bibfnamefont {C.-M.}\ \bibnamefont {Li}}\ and\ \bibinfo {author} {\bibfnamefont {Z.}~\bibnamefont {Quan}},\ }\bibfield  {title} {\bibinfo {title} {An efficient branch-and-bound algorithm based on maxsat for the maximum clique problem},\ }\bibfield  {booktitle} {\emph {\bibinfo {booktitle} {Proceedings of the AAAI Conference on Artificial Intelligence}},\ }\href@noop {} {\ \textbf {\bibinfo {volume} {24}},\ \bibinfo {pages} {128} (\bibinfo {year} {2010})}\BibitemShut {NoStop}%
\bibitem [{\citenamefont {Gray}\ and\ \citenamefont {Kourtis}(2021)}]{gray_hyperoptimized_2021}%
  \BibitemOpen
  \bibfield  {author} {\bibinfo {author} {\bibfnamefont {J.}~\bibnamefont {Gray}}\ and\ \bibinfo {author} {\bibfnamefont {S.}~\bibnamefont {Kourtis}},\ }\bibfield  {title} {\bibinfo {title} {Hyper-optimized tensor network contraction},\ }\href@noop {} {\bibfield  {journal} {\bibinfo  {journal} {Quantum}\ }\textbf {\bibinfo {volume} {5}},\ \bibinfo {pages} {410} (\bibinfo {year} {2021})}\BibitemShut {NoStop}%
\bibitem [{\citenamefont {Dudek}\ and\ \citenamefont {Vardi}(2020)}]{dudek_parallel_2021}%
  \BibitemOpen
  \bibfield  {author} {\bibinfo {author} {\bibfnamefont {J.~M.}\ \bibnamefont {Dudek}}\ and\ \bibinfo {author} {\bibfnamefont {M.~Y.}\ \bibnamefont {Vardi}},\ }\bibfield  {title} {\bibinfo {title} {Parallel weighted model counting with tensor networks},\ }\href@noop {} {\bibfield  {journal} {\bibinfo  {journal} {arXiv preprint arXiv:2006.15512}\ } (\bibinfo {year} {2020})}\BibitemShut {NoStop}%
\bibitem [{\citenamefont {Tamaki}(2019)}]{tamaki_positive-instance_2018}%
  \BibitemOpen
  \bibfield  {author} {\bibinfo {author} {\bibfnamefont {H.}~\bibnamefont {Tamaki}},\ }\bibfield  {title} {\bibinfo {title} {Positive-instance driven dynamic programming for treewidth},\ }\href@noop {} {\bibfield  {journal} {\bibinfo  {journal} {J. Comb. Optim.}\ }\textbf {\bibinfo {volume} {37}},\ \bibinfo {pages} {1283} (\bibinfo {year} {2019})}\BibitemShut {NoStop}%
\bibitem [{\citenamefont {Hamann}\ and\ \citenamefont {Strasser}(2018)}]{hamann_graph_2017}%
  \BibitemOpen
  \bibfield  {author} {\bibinfo {author} {\bibfnamefont {M.}~\bibnamefont {Hamann}}\ and\ \bibinfo {author} {\bibfnamefont {B.}~\bibnamefont {Strasser}},\ }\bibfield  {title} {\bibinfo {title} {Graph bisection with pareto optimization},\ }\href@noop {} {\bibfield  {journal} {\bibinfo  {journal} {J. Exp. Algorithmics}\ }\textbf {\bibinfo {volume} {23}},\ \bibinfo {pages} {1} (\bibinfo {year} {2018})}\BibitemShut {NoStop}%
\bibitem [{\citenamefont {Hicks}(2002)}]{hicks2002branchwidth}%
  \BibitemOpen
  \bibfield  {author} {\bibinfo {author} {\bibfnamefont {I.~V.}\ \bibnamefont {Hicks}},\ }\bibfield  {title} {\bibinfo {title} {Branchwidth heuristics},\ }\href@noop {} {\bibfield  {journal} {\bibinfo  {journal} {Congressus Numerantium}\ ,\ \bibinfo {pages} {31}} (\bibinfo {year} {2002})}\BibitemShut {NoStop}%
\bibitem [{\citenamefont {Nagy}\ \emph {et~al.}(2022)\citenamefont {Nagy}, \citenamefont {Paredes}, \citenamefont {Dudek}, \citenamefont {Due{\~n}as-Osorio},\ and\ \citenamefont {Vardi}}]{nagy_ising_2022}%
  \BibitemOpen
  \bibfield  {author} {\bibinfo {author} {\bibfnamefont {S.~A.}\ \bibnamefont {Nagy}}, \bibinfo {author} {\bibfnamefont {R.}~\bibnamefont {Paredes}}, \bibinfo {author} {\bibfnamefont {J.~M.}\ \bibnamefont {Dudek}}, \bibinfo {author} {\bibfnamefont {L.}~\bibnamefont {Due{\~n}as-Osorio}},\ and\ \bibinfo {author} {\bibfnamefont {M.~Y.}\ \bibnamefont {Vardi}},\ }\bibfield  {title} {\bibinfo {title} {Ising model partition function computation as a weighted counting problem},\ }\href@noop {} {\bibfield  {journal} {\bibinfo  {journal} {arXiv preprint arXiv:2212.12812}\ } (\bibinfo {year} {2022})}\BibitemShut {NoStop}%
\bibitem [{\citenamefont {Dechter}\ \emph {et~al.}(2002)\citenamefont {Dechter}, \citenamefont {Kask}, \citenamefont {Bin},\ and\ \citenamefont {Emek}}]{Dechter2002GeneratingRS}%
  \BibitemOpen
  \bibfield  {author} {\bibinfo {author} {\bibfnamefont {R.}~\bibnamefont {Dechter}}, \bibinfo {author} {\bibfnamefont {K.}~\bibnamefont {Kask}}, \bibinfo {author} {\bibfnamefont {E.}~\bibnamefont {Bin}},\ and\ \bibinfo {author} {\bibfnamefont {R.}~\bibnamefont {Emek}},\ }\bibfield  {title} {\bibinfo {title} {Generating random solutions for constraint satisfaction problems},\ }in\ \href {https://api.semanticscholar.org/CorpusID:3111334} {\emph {\bibinfo {booktitle} {AAAI/IAAI}}}\ (\bibinfo {year} {2002})\BibitemShut {NoStop}%
\end{thebibliography}%

\end{document}